\documentclass[pdflatex,sn-mathphys-num]{sn-jnl}

\usepackage{graphicx,subcaption}
\usepackage{multirow}%
\usepackage{amsmath,amssymb,amsfonts}%
\usepackage{amsthm}%
\usepackage{mathrsfs}%
\usepackage[title]{appendix}%
\usepackage{xcolor}%
\usepackage{textcomp}%
\usepackage{manyfoot}%
\usepackage{booktabs}%
\usepackage{algorithm}%
\usepackage{algorithmicx}%
\usepackage{algpseudocode}%
\usepackage{listings}%
\usepackage{soul}

\theoremstyle{thmstyleone}%
\newtheorem{theorem}{Theorem}
\theoremstyle{thmstyletwo}%
\newtheorem{assumption}{Assumption}%

\theoremstyle{thmstylethree}%

\begin{document}

\title[Article Title]{Hybrid State Estimation of Uncertain Nonlinear Dynamics Using Neural Processes}


\author*[1]{\fnm{Devin} \sur{Hunter}}\email{de700090@ucf.edu}

\author[1]{\fnm{Chinwendu} \sur{Enyioha}}\email{cenyioha@ucf.edu}
\equalcont{These authors contributed equally to this work.}

\affil*[1]{\orgdiv{Department of Electrical Engineering}, \orgname{University of Central Florida}, \orgaddress{\street{4000 Central Florida Blvd.}, \city{Orlando}, \postcode{32825}, \state{Florida}, \country{United States}}}


\abstract{Various neural network architectures are used in many of the state-of-the-art approaches for real-time nonlinear state estimation in dynamical systems. With the ever-increasing incorporation of these data-driven models into the estimation domain, models with reliable margins of error are required -- especially for safety-critical applications. This paper discusses a novel hybrid, data-driven state estimation approach based on the physics-informed attentive neural process (PI-AttNP), a model-informed extension of the attentive neural process (AttNP). We augment this estimation approach with the regression-based split conformal prediction (CP) framework to obtain quantified model uncertainty with probabilistic guarantees. After presenting the algorithm in a generic form, we validate its performance in the task of grey-box state estimation of a simulated under-actuated six-degree-of-freedom quadrotor with multimodal Gaussian sensor noise and several external perturbations typical to quadrotors. Further, we compare outcomes with state-of-the-art data-driven methods, which provide significant evidence of the physics-informed neural process as a viable novel approach for model-driven estimation.}




\maketitle

\section{Introduction}\label{sec:intro}

The modeling of dynamical systems under stochastic disturbances and nonlinearities remains a major challenge in robotics, aerospace, and process control. Deep learning-based estimators have recently gained attention as a technique for addressing these challenges. Examples include Deep Kalman Filters (DKFs) such as KalmanNet \cite{revach2022kalmannet} and Deep Variational Bayes Filters (DVBF) \cite{karl2017dvbf}, which embed neural dynamics into recursive Bayesian frameworks. Additional deep learning-based estimators are Physics-Informed Neural Networks (PINNs) enhanced with Monte Carlo dropout such as the PINN-UKF \cite{Curto2024Pinn} and Damper-B-PINN \cite{zeng2025pinn}, which embed known dynamics into neural architectures to enforce consistency with system equations and enable uncertainty estimation.

Despite these advances, existing methods face specific challenges. For example, DKFs typically rely on recursive structures that introduce inference latency, while Gaussian assumptions limit the expressiveness of their uncertainty modeling \cite{revach2022kalmannet}. Additionally, PINNs often assume that utilized governing equations fully characterize the dynamics of a system while also often using approximate uncertainty modeling techniques such as MC dropout \cite{Curto2024Pinn}. To address these limitations, recent studies have incorporated physical knowledge into learning architectures. Pertaining to issues of mentioned PINN methods, as previously stated, these models often rely on dropout sampling rather than full probabilistic modeling, which limits scalability and precision under irregular sampling. Other uncertainty quantification (UQ) techniques such as deep ensembles, Bayesian neural networks and variational latent models are known to introduce considerable computational overhead or result in suboptimal uncertainty estimates \cite{fort2020deepensembleslosslandscape,kristiadi2020bayesianjustbitfixes}.

To address these drawbacks, we introduce a hybrid framework that couples our understanding of physical models with the advantages of the Attentive Neural Process (AttNP). Our novel approach, called the Physics Informed Attentive Neural Process (PI-AttNP), sidesteps computation-heavy inference and instead models distributions over functions directly, allowing for scalable, non-Gaussian uncertainty quantification via context-aware attention wrapped with the statistical guarantees provided from split conformal predictions. Its physics-informed priors act as structural constraints during learning, mitigating the reliance on purely data-driven inference and hand-tuned noise models while maintaining fast inference and robustness under sparse, noisy observations.

Our deep learning-based estimation algorithm performs fast inference in real-time using irregularly sampled and noisy data. This is accomplished using the powerful capabilities of PI-AttNP in scalable, predictive sequential modeling that  accurately captures the stochastic processes inherent in the state estimation problem \cite{kim2019attentiveneuralprocesses,garnelo2018neuralprocesses}. This paper specifically presents:
\begin{itemize}
\item a PI-AttNP architecture for state estimation of a general nonlinear dynamical system under unknown external perturbations and sensor noise;
\item guarantees on the error and uncertainty bound based on marginal conformal prediction; and
\item direct experimental validation and comparative analysis of our proposed method against several state-of-the-art hybrid methods, including the physics-informed neural net (PINN) and deep Kalman filter (DKF) algorithm on a simulated quadrotor.
\end{itemize}
We show that our PI-AttNP approach provides robust and accurate estimation through implicit learning of sensor noise distributions and other environmental variations directly from the data.

The remainder of this paper is structured as follows. In Section \ref{sec:problem-statement}, we formally state the problem and follow in Section \ref{sec:lit-review} with a brief overview of state-of-the-art approaches with which we contrast our approach. Sections \ref{sec:method} and \ref{sec:results} present the approach and outcomes, respectively. We make concluding remarks in Section \ref{sec:conclude+future}.

\section{Problem Statement}\label{sec:problem-statement}

In this study, the objective is to learn the model of a vector field of coupled, nonlinear equations $f$ for a generic robotic system given in the following discrete-time form:
\begin{equation}\label{eqn:dt-model}
  \begin{aligned}
    x_{k+1} &= f\left(x_k, u_k,\xi_k,\Delta x_{k},\Delta k\right) \\
    y_{k} &= h(x_{k})
  \end{aligned}
\end{equation}
where $x_k \in \mathbb{R}^{n}$ are the system states, $u_k \in \mathbb{R}^{m}$ are system controls, $y_{k}\in\mathbb{R}^{o}$ are the observed states and $h:\mathbb{R}^{n}\rightarrow\mathbb{R}^{o}$ is an arbitrary, nonlinear measurement model. Additionally, $\xi_k\in\mathbb{R}^{o}$ represents a time-varying multimodal noise vector with unknown additive/multiplicative characteristics that perturbs the observed states of the system, $\Delta x_{k}\in\mathbb{R}^{n}$ represents external disturbances that affect the system dynamics, and $\Delta k$ represents the user-defined, non-uniform time step between states $x_{k}$ and $x_{k+1}$.

To accomplish this nonlinear estimation, our proposed PI-AttNP learns a surrogate model $\hat{f}_{\Gamma}(x_{k},u_{k},\Delta k):\mathbb{R}^{n}\times\mathbb{R}^{m}\times\mathbb{R} \rightarrow \mathbb{R}^{n}$ that represents the approximate dynamics of a general nonlinear system $f: \mathbb{R}^{n}\times\mathbb{R}^{m}\times\mathbb{R}^{n}\times\mathbb{R}^{n}\times\mathbb{R} \rightarrow \mathbb{R}^{n}$ given model parameters $\Gamma$. With the inclusion of a simplified physics model $g(x_{k},u_{k},\Delta k): \mathbb{R}^{n}\times\mathbb{R}^{m}\times\mathbb{R}\rightarrow\mathbb{R}^{n}$ that does not consider how sensor noise nor environmental disturbances affect the dynamics of the system, the learned dynamics $\hat{f}_{\Gamma}$ can be implicitly represented as:
\begin{equation}
    \hat{f}_{\Gamma} = g(x_k,u_k,\Delta k) + NN(x_k,u_k,\Delta k|\Gamma),
\end{equation}
where $ NN(\cdot) $ represents the learned neural network residual of $f$ that incorporates noise and other dynamic effects. The inclusion of $g(\cdot)$ enhances the AttNP's predictive capabilities with physics-informed priors that loosely describe the dynamics of the perturbed system. Despite its inclusion, it is important to note that the nature of the learning convergence of our model of $f$ does not require prior knowledge of the characteristics of sensor noise nor external disturbances to learn a wide range of system dynamics.

\section{Background and Literature Review}\label{sec:lit-review}
The problem of state estimation has been extensively studied over the past several decades, with a wide range of techniques developed to infer unobservable system states from noisy measurements. Early approaches predominantly relied on model-based, or white-box, techniques that assumed an accurate mathematical description of the system dynamics. Among the most notable are the Kalman Filter (KF) for linear Gaussian systems \cite{Kalman1960}, adaptive filtering extensions that update state beliefs based on online parameter estimation \cite{Isaksson1987AdpaptiveKF}, and Moving Horizon State Estimation (MHSE), which formulates state estimation as an optimization problem over a finite time horizon \cite{LIU2013MHSE}. These approaches have proven effective for systems where the underlying dynamics are well-understood and sensors provide reliable, observable information.

However, in many real-world applications, particularly those involving complex, nonlinear dynamics, non-Gaussian noise, and unmodeled exogenous disturbances, traditional model-based methods struggle to maintain reliable performance. Factors such as stochastic environmental perturbations, modeling errors, and sensor limitations often introduce modeling inaccuracies that can severely degrade estimation performance \cite{jin2021new,BAI2023100909}.

In response to these limitations, purely data-driven approaches have emerged as an alternative, bypassing the need for explicit dynamical models. Techniques such as recursive least squares (RLS) \cite{wevj1509RLS}, support vector machines (SVM) \cite{Sanchez2004SVM}, and early forms of artificial neural networks (ANNs) \cite{Kumar2011ANN,Wilson1997ANN} have been applied to state estimation tasks, demonstrating the ability to approximate complex system dynamics directly from measurement data. Despite their flexibility, purely data-driven models often lack the inherent physical structure present in model-based methods, leading to predictions that may violate fundamental system constraints or yield physically implausible results \cite{jin2021new,feng2023review}.

To address these shortcomings, hybrid, or grey-box, approaches have gained significant traction by fusing the structured insights of model-based methods with the expressive power of machine learning. Two prominent examples of such hybrid approaches are the Deep Kalman Filter (DKF) and Physics-Informed Neural Networks (PINNs).
The Deep Kalman Filter represents one of the earliest and most influential attempts to extend classical Bayesian filtering by incorporating deep learning into the estimation process \cite{krishnan2015DKF}. Recent advances have significantly expanded the DKF framework, introducing architectural innovations to improve expressiveness, scalability, and robustness. Notably, works such as deep variational Bayes filters (DVBFs)\cite{karl2017deepvariationalbayesfilters} and probabilistic recurrent state-space models (PR-SSMs)\cite{Yuan2023SSM} have enabled more flexible modeling of latent system dynamics and uncertainty. Other state-of-the-art extensions have incorporated attention mechanisms, structured latent representations, and multi-modal uncertainty modeling, making DKFs highly effective for capturing complex, nonlinear, and stochastic dynamics in partially observed environments \cite{goel2024transformerrepresentkalmanfilter}.

Despite these advancements, DKFs continue to rely on inner recursive filtering algorithms, which introduces added computation in real-time scenarios where methods with lower inference latency may be more desirable. Moreover, while DKFs can approximate time-varying process and measurement noise, errors in their estimation due to their heteroscedastic nature can propagate over towards the construction of state covariance matrix $P_{k}$, making them susceptible to generating implausible state estimates, especially in poorly observed or highly uncertain regimes. While multiple recent efforts have investigated learned covariance augmentations to minimize this approximation error, a less expensive option is desirable.

Physics-Informed Neural Networks have recently emerged as a leading paradigm for blending physical modeling with deep learning, offering a systematic way to incorporate governing equations into neural network training. While early PINN approaches focused on solving partial differential equations (PDEs) \cite{karniadakis2021physics}, recent state-of-the-art work has expanded their use to real-time state estimation, system identification, and control for complex, nonlinear, and partially known systems \cite{ANTONELO2024PINN}. Innovations such as adaptive weighting of physics residuals, physics-informed recurrent networks, and hybrid PINN-Gaussian Process models have enhanced the applicability of PINNs to challenging real-world problems, including fluid dynamics, robotics, and biological systems \cite{alla2025pinn,ngo2023physicsinformedgraphicalneuralnetwork,TARTAKOVSKY2023PINN}.

Despite these advances, PINNs face several persistent challenges when applied to sequential state estimation tasks. Many PINN implementations assume fully or partially known governing equations, limiting their use in scenarios where the system dynamics are highly uncertain or partially observed \cite{karniadakis2021physics}. In addition, uncertainty quantification in PINNs is often approximate, for example via Monte Carlo dropout, and does not capture the full epistemic and aleatoric uncertainty present in complex, noisy environments\cite{Curto2024Pinn}. 

Neural Processes represent a family of deep generative models that learn distributions over functions using meta-learning principles \cite{dubois2020npf}. Inspired by Gaussian Processes, NPs provide stochastic, uncertainty-aware predictions without requiring the explicit definition of a kernel function or suffering from scalability limitations typical of GPs \cite{dubois2020npf,garnelo2018neuralprocesses}. NPs have demonstrated success in tasks such as irregular time series forecasting and image completion, where fast adaptation to new tasks and uncertainty quantification are critical \cite{dubois2020npf}.

Despite these strengths, original NP formulations exhibit underfitting near contextual data points, primarily due to the order-invariant aggregation mechanism used to summarize context inputs \cite{kim2019attentiveneuralprocesses}. This limitation is particularly problematic for state estimation tasks, where accurate exploitation of recent measurements is essential for reliable future state predictions.

To overcome this, the Attentive Neural Process introduces a cross-attention mechanism, enabling the model to selectively focus on informative regions of the input space \cite{kim2019attentiveneuralprocesses, dubois2020npf}. This significantly enhances generalization to unseen contexts and reduces underfitting. Building on this, the PI-AttNP incorporates known system dynamics as an informed prior over the NP predictive distribution. This hybrid design unites the fast adaptation and uncertainty modeling of NPs with the physical consistency offered by embedded dynamic models while avoiding the sequential computational bottleneck of previous NP models.

Alternative state estimation methods such as particle filters and graph neural network-based estimators have also shown promise in recent years \cite{wang2017particle,ngo2023physicsinformedgraphicalneuralnetwork}. These techniques offer increased flexibility for modeling complex and nonlinear dynamics, especially in partially observed systems. However, they often require extensive labeled data, can be computationally demanding, or lack explicit mechanisms for enforcing physical consistency.

The PI-AttNP offers a novel, scalable, and physically consistent approach to state estimation that addresses key limitations of existing techniques, particularly in real-world environments where dynamics are partially known, noisy, and subject to exogenous disturbances. It is important to note that the PI-AttNP shares structural similarities to the multi-fidelity neural process (MFNP) that leverages combinations of low-fidelity surrogate models in the NP decoder to make inference on high-fidelity ground truth data \cite{Wu2022MFHNP,niu2024multifidelityresidualneuralprocesses}. However, our novel contrast to this approach is the explicit use of a low-fidelity (simple) physics model in approximating the high-fidelity (complex) dynamics needed for estimation.  

\section{Methodology}\label{sec:method}
The objective is to estimate the nonlinear dynamics perturbed by external disturbance $\Delta x$ and multimodal sensor noise $\xi$ drawn from a distribution with unknown parameters using the Physics-Informed Attentive Neural Process (PI-AttNP). To understand the structure of this proposed architecture, observe the computational diagram of the original AttNP with the inclusion of $g(\cdot)$ (i.e. the PI-AttNP) illustrated in Figure \ref{fig:attnp-forward}.

\subsection{Model Forward Pass Discussion}

\begin{figure}
    \centering
    \includegraphics[scale=0.4]{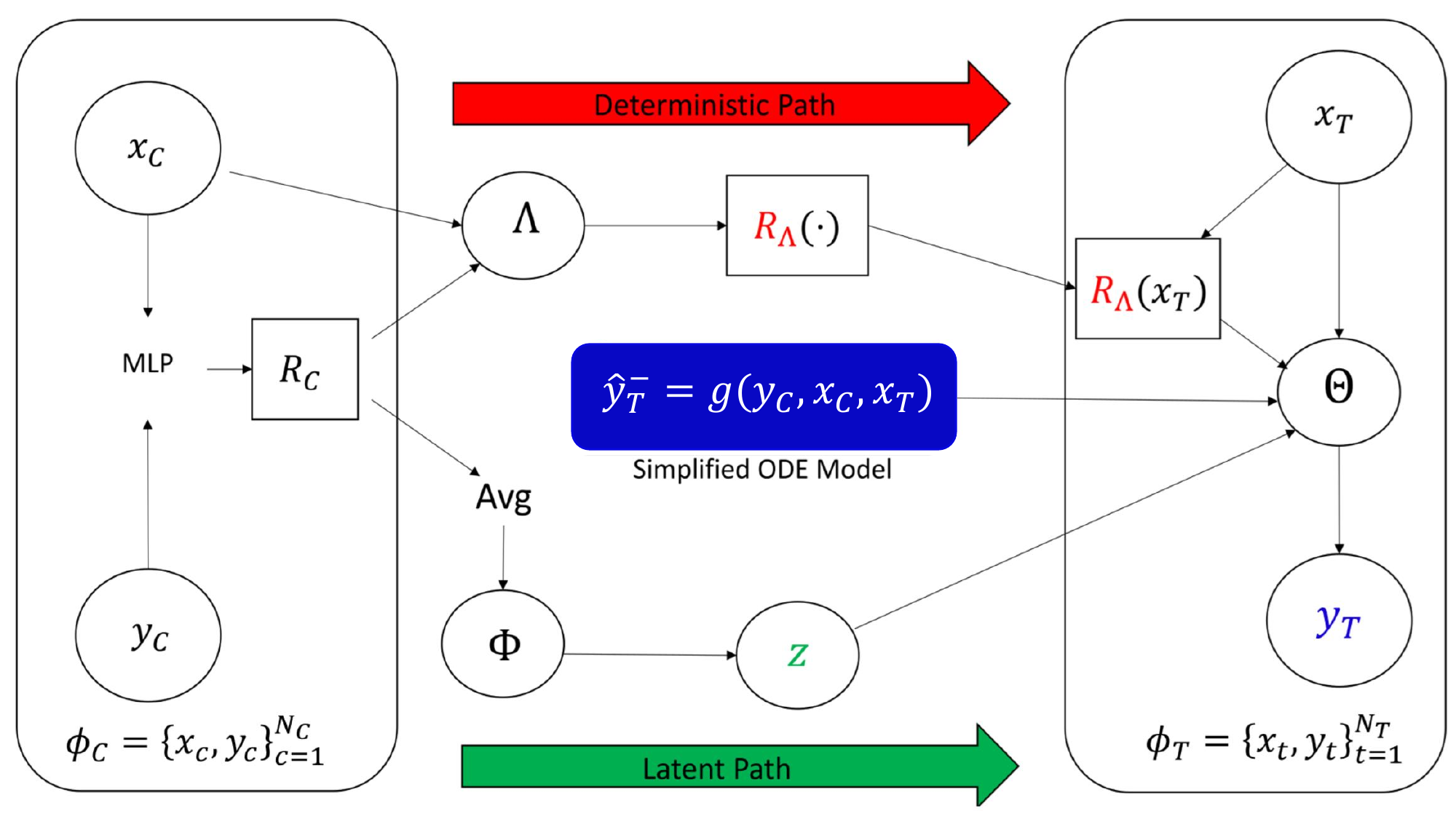}
    \caption{Computational Graph of PI-AttNP's Forward Pass}
    \label{fig:attnp-forward}
\end{figure}

As seen in Figure \ref{fig:attnp-forward}, the PI-AttNP maps contextual sets $\phi_{C} = \left\{x_{c}, y_{c}\right\}_{c=1}^{N_{C}}$ to a global representation $R_{C}$ through an initial embedding via a Multi-Layer Perceptron (MLP). Within the estimation domain, it is noted that contextual sets $\phi_{C}$ can represent current dynamic information propagating from the system, such as a control action $u_{k}$ ($x_{C}$) and previous state estimate $\hat{x}_{k}$ ($y_{C}$).

After dynamic set $\phi_{C}$ is transformed into $R_{C}$, it is sent through split deterministic and latent paths within the forward pass of the network. In the latent path, $R_{C}$ is, first, sent through an averaging operation $\textbf{Avg}$ across individual representations $r_{c}\in R_{C}$ derived from each $\{x_{c},y_{c}\}\in \phi_{C}$. This order-invariant computation is performed to ensure that $\phi_{C}$ maintains the required exchangability condition needed to parameterize the latent variable $z$. After this step, averaged $\phi_{C}$ is used as input to a latent encoder MLP $\Phi(\cdot)$ that is used in the variational approximation of prior distribution $p(z|\phi_{C})$ and posterior distribution $p(z|\phi_{T})$; where $\phi_{T} = \left\{x_{t}, y_{t}\right\}_{t=1}^{N_{T}}$ denotes the target set. Within the estimation domain, the target set $\phi_{T}$ captures future dynamic information such as time step $\Delta k$ ($x_{T}$) and future state estimates $\hat{x}_{k+1}$ ($y_{T}$).

Through the deterministic path, $R_{C}$ is transformed into a \emph{context-aware} attention matrix $R_{\Lambda}(x_{C},y_{C},x_{T})$ using a deterministic encoder MLP, $\Lambda(\cdot)$, which possesses a cross attention mechanism. Our AttNP approach utilizes multi-headed attention within the cross attention mechanism found in $\Lambda$, which computes parallel scaled dot-products of contextual inputs $x_{C}$ (keys), target inputs $x_{T}$ (queries), and global representation $R_{C}$ (values). Our approach builds on the results in \cite{kim2019attentiveneuralprocesses}, where the benefits of utilizing eight parallel heads to attain maximum learning efficiency were highlighted. Furthermore, our approach uses self-attention between context pairs $\{x_{C},y_{C}\}$ to better model how aspects of control $u_{k}$ affect the current state estimate $\hat{x}_{k}$ from a transitional dynamics perspective \cite{dubois2020npf}.

Our proposed architecture utilizes a decoder MLP, $\Theta(\cdot)$, to map a set of queried inputs $x_{T}$, attention matrix $R_{\Lambda}$, latent sample $z\sim\mathcal{Z}$, and prior estimated states $\hat{y}_{T}^{-}\in\mathbb{R}^{n}$ from $g(\cdot)$ (See Problem Statement \ref{sec:problem-statement}) to a predictive distribution over queried outputs $p_{\Theta}(y_T | x_{T}, R_{\Lambda},z,\hat{y}_{T}^{-})$ \cite{dubois2020npf, kim2019attentiveneuralprocesses}. Based on these decoder inputs, a goal is for our model to infer $y_{T}$ using only the given contextual information found in $\phi_{C}$, $x_{T}$, and prior estimate $\hat{y}_{T}^{-}$ computed from simplified dynamics model $g(\cdot)$. For a general estimation task, the decoder of the AttNP will output posterior state estimates $\hat{y}_{T}^{+} \sim \rho$ where $\rho$ can be any arbitrary state probability distribution. To capture statistical guarantees from model uncertainty, note that predicted variance $\hat{\sigma}^{2}\in \mathbb{R}^{n}$ generated from $p_{\Theta}$ are linearly scaled with quantile vector $q_{\alpha}\in \mathbb{R}^{n}$ computed from parallel marginal conformal predictions. To further understand our conformal prediction approach and the theoretical guarantees on prediction uncertainty and error bounds it provides, observe the following subsection \ref{subsec:cp}.

It should also be noted that given the roots of PI-AttNP in probabilistic modeling from GPs\cite{dubois2020npf}, we can define context and target set sizes $N_{C},N_{T}$ to be arbitrarily set. Within our one-step estimation application, we set $N_{C}=N_{T}=1$ to ensure that only the most recent sensor measurement is used to infer the next states within a Markovian framework. This was done to ensure that this estimation approach can be compared directly with other baseline approaches that operate based on Markovian assumptions \cite{BAI2023100909,feng2023review,wang2017particle}. However, it should be explicitly noted that our PI-AttNP can also be applied to non-Markovian systems given the proper supply of contexts $N_{C}$ to model needed temporal relationships. Furthermore, based on this flexible context/target setup, it is also possible for the PI-AttNP to predict future state trajectories (i.e. $\hat{x}_{k+1},\hat{x}_{k+2},\dots,\hat{x}_{k+N}$) rather than the single-step prediction framework that we employ currently.  

\subsection{Model Optimization Algorithm}

To understand how the PI-AttNP incorporates the discussed AttNP algorithm with the addition of simplified model $g(\cdot)$ in Equation \eqref{sec:problem-statement}, observe Algorithm \ref{alg:attnp-training-alg}. Within the algorithm, we note that $\widetilde{y}_{C}$ is the version of $y_{C}$ perturbed by noise vector $\xi$, $\epsilon_{\text{done}}$ is a model training stopping criterion based on testing loss $\mathcal{L}_{\text{test}}$, $\mathcal{D}_{\text{meta}}$ represents the meta dataset used to train our models, and $\mathcal{D}_j$ can be a singular task set (sample) or a mini-batch. In our implementation, we trained the AttNP using the mini-batch approach (with $\text{size}(\mathcal{D}_{j}) = 1000,\; \text{size}(\mathcal{D}_{\text{meta}}) = 2\times 10^5$) to avoid noisy gradients and achieve stable convergence. Concerning Adam optimization parameters (appearing in line 23), we utilize a learning rate $lr = 1\times 10^{-3}$ and weight decay regularization value $\lambda = 1 \times 10^{-6}$. To enforce training consistency between all trained models, we choose the same training hyperparameter values for all deep learning-based baselines mentioned in \ref{subsec:baselines} as well. Regarding model $g(\cdot)$ in Equation \eqref{sec:problem-statement}, prior estimated next states $\hat{y}_{T}^{-}$ are incorporated into the parameterized predictive distribution $p_{\Theta}(y_{T}|x_{T},z,R_{\Lambda},\hat{y}_{T}^{-})$. Note that $\hat{y}_{T}^{-}$ derived from $g(\cdot)$ is computed based on relevant dynamic information found from $x_{C},y_{C},x_{T}$. It is imperative to understand that surrogate distributions $q_{\Phi}(\cdot)$ and predictive distribution $p_{\Theta}(\cdot)$ are modeled as Gaussians due to their flexibility in approximating arbitrary distributions.

 To train this model, we solve the following variational lower-bound (ELBO) optimization:
\[ p_{\Theta}^{*}, q_{\Phi}^{*}, R_{\Lambda}^{*} = \arg\max_{\Gamma^{*}} \mathcal{L}(p_{\Theta}, q_{\Phi},R_{\Lambda}),
\]
    subject to
\begin{align}
\label{eq:optimization}
    \text{log}\;p(y_{T}|x_{T}, \phi_{C}) &\geq \mathcal{L}(p_{\Theta},q_{\Phi},R_{\Lambda}) \nonumber \\
    &= \mathbb{E}_{z \sim q_{\Phi}}\text{log}\left[p_{\Theta}
(y_{T}|x_{T}, R_{\Lambda}, z,\hat{y}_{T}^{-})\right] \nonumber \\
       &- \mathbb{E}_{z \sim q_{\Phi}}\text{log}\left[\frac{q_{\Phi}(z|\phi_{T})}{q_{\Phi}(z|\phi_{C})}\right], 
\end{align}
%
%
where $\Gamma^{*} = \left\{\Theta^{*},\Phi^{*},\Lambda^{*}\right\}$ and the constraint in Equation \eqref{eq:optimization} is the KL-divergence between the approximated latent posterior $q_{\Phi}(z|\phi_{T})$ and prior $q_{\Phi}(z|\phi_{C})$, expressed as 
\[ D_{KL}(q_{\Phi}(z|\phi_{T}) || q_{\Phi}(z|\phi_{C})).\] 

The key intuition of the above optimization is that we desire to obtain model parameters $\Gamma^*$ that maximize the marginal likelihood $\log p(y_{T}|x_{T},\phi_{C})$. Since latent sample $z$ is introduced as an additional conditioning variable in the decoder's parameterized distribution (i.e. $p_{\Theta}
(y_{T}|x_{T}, R_{\Lambda}, z)$), a marginalization of $z$ is required to obtain the target distribution. However, this marginalization is often intractable due to the high dimensionality of $z$ in addition to the lack of knowledge pertaining all latent samples $z\in\mathcal{Z}$. Therefore, an approximate variational method is employed to \emph{indirectly} optimize the intractable left side of the inequality found in \ref{eq:optimization} by maximizing computationally-feasible right side of the inequality. In Algorithm \ref{alg:attnp-training-alg}, this variational loss is computed for both $\mathcal{L}_{\text{train}}$ and $\mathcal{L}_{\text{test}}$ in lines 22 and 30. 

\begin{algorithm}
\caption{Offline Training Procedure for PI-AttNP}
\label{alg:attnp-training-alg}
\begin{algorithmic}[1]
\Require $\Gamma_{0} = \{\Phi_{0},\;\Theta_{0},\;\Lambda_{0}\},\;\mathcal{D}_{\text{meta}} = \{\mathcal{D}_{\text{train}}, \mathcal{D}_{\text{test}}\},\;g(\cdot)$
\Ensure $\mathcal{D}_{\text{train}} \geq\; 0.6 \mathcal{D}_{\text{meta}}$
\While{$\mathcal{L}_{\text{test}} > \epsilon_{\text{done}}$}
\For{$\mathcal{D}_{j}$ in $\mathcal{D}_{\text{train}}$}
\State \textbf{Parsing Context and Target Sets} 
\State $\; x_{C}, \widetilde{y}_{C}, x_{T}, y_{T} \gets \mathcal{D}_{j}$ \\
\State $\textbf{Computing Latent Prior and Posterior}$
\State $\; q_{\Phi}(z|\phi_{C}) \gets \Phi_{i}(x_{C},\widetilde{y}_{C})$
\State $\; q_{\Phi}(z|\phi_{T}) \gets \Phi_{i}(x_{T},y_{T})$ \\
\State $\textbf{Sampling Latent Variable from Posterior}$
\State $\textbf{and Computing Attention Matrix}$
\State $\; z \sim q_{\Phi}(z|\phi_{T})$
\State $\; R_{\Lambda} \gets \Lambda_{i}(x_{C},\widetilde{y}_{C},x_{T})$ \\
\State $\textbf{Computing Apriori Estimate over}$
\State $\textbf{Next States}$
\State $\hat{y}_{T}^{-} \gets g(\widetilde{y}_{C},x_{C},x_{T})$ \\
\State $\textbf{Computing Predictive Distribution over}\; y_{T}$
\State $\textbf{and Training Loss to Optimize AttNP}$
\State $p_{\Theta} \gets \Theta_{i}(x_{T},z,R_{\Lambda},\hat{y}_{T}^{-})$
\State $\mathcal{L}_{\text{train}} \gets \text{ELBO}(p_{\Theta},y_{T},q_{\Phi}(z|\phi_{C}),q_{\Phi}(z|\phi_{T}))$
\State $\Gamma_{i+1} \gets \text{Adam}(\mathcal{L}_{\text{train}},\Gamma_{i})$
\EndFor
\State $\textbf{Computing}\;\mathcal{L}_{\text{test}}\;\textbf{Using Sample(s) from}\;\mathcal{D}_{\text{test}}$
\State $\mathcal{D}_{j} \sim \mathcal{D}_{\text{test}}$
\State $\text{Repeat lines (4) - (8) with}\; \mathcal{D}_j$
\State $z \sim q_{\Phi}(z|\phi_{C})$
\State $\text{Repeat lines (13) - (17)}$
\State $\mathcal{L}_{\text{test}} \gets \text{ELBO}(p_{\Theta},y_{T},q_{\Phi}(z|\phi_{C}), q_{\Phi}(z|\phi_{T}))$
\EndWhile
\State $\textbf{return}\;\;\Gamma_{\text{end}}\approx \;\Gamma^{*}$
\end{algorithmic}
\end{algorithm}

\begin{algorithm}
\caption{PI-AttNP Recursive Inference Algorithm}
\label{alg:recursive-estimation-alg}
\begin{algorithmic}[1]
\Require $\hat{f}_{\Gamma}(\cdot),\; g(\cdot),\;\mathcal{D}_{k},\;q_{\alpha}$
\State $\textbf{Parsing NP Query from Current Task Set $\mathcal{D}_{k}$}$
\State $\; x_{C}, \widetilde{y}_{C}, x_{T} \gets \mathcal{D}_{k}$ 
\State $\left(\textbf{Note that: }\widetilde{y}_{C}=\hat{x}_{k},x_{C}=u_{k}, x_{T}=\Delta k\right)$ \\
\State $\textbf{Computing Apriori Estimate over}$
\State $\textbf{Next States}$
\State $\hat{y}_{T}^{-} \gets g(\widetilde{y}_{C},x_{C},x_{T})$ \\
\State $\textbf{Computing Predictive Distribution over}\; y_{T}$
\State $\hat{y}_{T}^{+}, \hat{\sigma}_{T}^{2} \gets \hat{f}_{\Gamma}(x_{C},\widetilde{y}_{C},x_{T},\hat{y}_{T}^{-})$ \\
\State $\textbf{Scale Model Uncertainties via CP Quantile}$
\State $\hat{\sigma}_{T}^{+} \gets q_{\alpha} \cdot \sqrt{\hat{\sigma}_{T}^{2}}$ \\
\State $\textbf{Compute Uncertainty Weight }\;\beta$
\State $\beta_{k} = 1 \;/\;(1+\hat{\sigma}_{\text{obs}}^{+}) \in \mathbb{R}^{o}$ \\
\State $\textbf{Update Next Context} \;\widetilde{y}_{C}\; \textbf{and Task Set} \;\mathcal{D}_{k+1}$
\State $\widetilde{y}_{C} \gets \beta_{k} \cdot \hat{y}_{T}^{+} + (1-\beta_{k}) \cdot y_{k}$
\State $x_{C}\gets u_{k+1}$
\State $x_{T}\gets \Delta k$
\State $\mathcal{D}_{k+1} \gets x_{C},\widetilde{y}_{C},x_{T}$ \\
\State $\textbf{return}\; \hat{y}_{T}^{+},\;\hat{\sigma}_{T}^{+},\;\mathcal{D}_{k+1}$

\end{algorithmic}
\end{algorithm}

\subsection{Recursive Estimation Algorithm Discussion} \label{subsec:recursive-alg}
In Algorithm \ref{alg:recursive-estimation-alg}, one can observe how our PI-AttNP-based recursive state estimation algorithm operates in sequential prediction tasks. Regarding input arguments, note that $\mathcal{D}_{k}$ represents the previous task set that contains dynamic information ($x_{C},\widetilde{y}_{C},x_{T}$) needed to predict the next immediate states $y_{T}$. The conformal quantile parameter $q_\alpha$ is used to linearly scale the model uncertainty standard deviations $\hat{\sigma}_{T}$ in order to obtain probabilistic guarantees for ground-truth predictions and error bounds for each state in $y_{T}$. Recall that $\hat{f}_{\Gamma}$ represents the learned PI-AttNP based dynamics model referenced previously \ref{sec:problem-statement}. In line 19, when the noisy context signal $\widetilde{y}_{C}$ is updated, it does so using a weighted average between the posterior average $\hat{y}_{T}^{+}$ and the most recent noise-perturbed sensor measurements $y_{k}$. This action is done to provide an error correction mechanism to mitigate the accumulation of prediction error over time. Note that since $\widetilde{y}_{C}$ represents the previous state vector $\hat{x}_{k}$, we observe that only observable states of $\hat{y}_{T}^{+}$ are fused with noisy observations $y_{k}$. This weighted computation is controlled through a time-varying, model uncertainty-based weighting term $\beta_k\in[0,1]$ computed in line 16 where the \emph{observed} state components of the conformal quantile-scaled PI-AttNP uncertainty $\hat{\sigma}_{\text{obs}}^{+}\in\mathbb{R}^{o}$ are utilized in its computation. Through this weighting scheme, the model adaptively varies the contributions of observed state predictions and true observations throughout an estimation task. Since $\hat{\sigma}_{\text{obs}}^{+}$ now provides statistical robustness from the marginal CP approach, the PI-AttNP is expected to learn the transitional relationship between previous fused estimate $\widetilde{y}_{C}$, updated prediction $\hat{y}_{T}^{+}$, and ground truth next states $y_{T}$ at each recursive iteration.  

\subsection{Conformal Prediction for Calibrated Uncertainty \newline Guarantees}\label{subsec:cp}

To quantify the uncertainty in our PI-AttNP-based state estimates and provide probabilistic guarantees, we integrate a marginal conformal prediction (CP) framework. This approach constructs data-driven prediction intervals for each state dimension, ensuring that the true future state \( y_T \in \mathbb{R}^n \) lies within these intervals with user-defined confidence \( 1 - \alpha \), without making strong assumptions on the underlying data distribution.

\subsubsection{Conformal Set Construction}

Let \( \hat{y}^{(j)}_T \) and \( \hat{\sigma}^{(j)}_T \) denote the predicted mean and standard deviation, respectively, for state dimension \( j \in \{1, \dots, n\} \) from the PI-AttNP $\hat{f}$ model. We define the score function $s(x_{i},y_{i}) = s_{i}$ as normalized squared residuals for each calibration sample \( (x_i, y_i) \) and state dimension \( j \) in the following form:

\begin{equation}\label{eq:conformity-score}
s_i^{(j)} = \frac{\left(y_i^{(j)} - \hat{y}_i^{(j)}\right)^{2}}{\hat{\sigma}_i^{(j)}},
\end{equation}
where $\hat{y}_{i},\; \hat{\sigma}_{i}=\hat{f}(x_{i})$. Given a calibration dataset \( \mathcal{D}_{\text{cal}} \) of size $N$, we compute the empirical quantile \( q_\alpha^{(j)} \) of the scores \( \{s_1^{(j)}, \dots, s_N^{(j)}\} \) such that:

\[
q_\alpha^{(j)} = \text{Quantile}_{1-\alpha} \left( \{ s_i^{(j)} \}_{i=1}^N \cup \{+\infty\} \right).
\]
The resulting conformal prediction interval $C_{\alpha}^{(j)}(x_{T})$ for the test-time state prediction in dimension \( j \) is:

\[
\mathcal{C}_\alpha^{(j)}(x_{T}) = \left[ \hat{y}^{(j)}_T - q_\alpha^{(j)} \cdot \hat{\sigma}^{(j)}_T, \,\;\; \hat{y}^{(j)}_T + q_\alpha^{(j)} \cdot \hat{\sigma}^{(j)}_T \right].
\]
\subsubsection{Coverage Analyses and Estimation Error Guarantees}
To establish the guaranteed error bounds needed for our CP-based estimation approach, we assume the following: 

\begin{assumption} The conformity scores $ \{s_i^{(j)}\}_{i=1}^{N} $ computed on the calibration set as shown in Equation \eqref{eq:conformity-score} and the test score \( s_{N+1}^{(j)} \) are jointly exchangeable and drawn from a continuous distribution.
\end{assumption}
The technical assumptions above are made to ensure that the scores can be strictly ordered with probability 1.

\begin{theorem}\label{thm:coverage}
    Under the assumption that the calibration scores \( \{s_1^{(j)}, s_2^{(j)}, \dots, s_N^{(j)} \} \) and the test score \( s_{N+1}^{(j)} \) are exchangeable and drawn from a continuous distribution, the marginal split conformal framework guarantees the following tight probabilistic coverage for each state dimension \( j \in \{1, \dots, n\} \):
    \[
        1 - \alpha \leq \mathbb{P}\left( y_{N+1}^{(j)} \in \mathcal{C}_\alpha^{(j)}(x_{N+1}) \right) \leq 1 - \alpha + \frac{1}{N+1}
    \]
\end{theorem}

\begin{proof}
We establish the proof of Theorem \ref{thm:coverage} in phases noted with bold font below, beginning with the quantile construction and ending with the upper bound on the error.

\noindent \textbf{Quantile construction:} For a given state dimension \( j \), define the \((1 - \alpha)\)-quantile as:
\[
    q_\alpha^{(j)} = s^{(j)}_{\left\lceil (N+1)(1 - \alpha) \right\rceil},
\]
where \( s^{(j)}_{(1)} \leq s^{(j)}_{(2)} \leq \dots \leq s^{(j)}_{(N)} \) denotes the ordered calibration scores, and \( q_\alpha^{(j)} \) is the \( \left\lceil (N+1)(1 - \alpha) \right\rceil \)-th smallest value in the ordered set \( \{s_1^{(j)}, \dots, s_N^{(j)}, s_{N+1}^{(j)}\} \).

\noindent \textbf{Uniform rank argument:} Under exchangeability, the rank of the test score \( s_{N+1}^{(j)} \) is uniformly distributed over the \( N + 1 \) possible positions. Therefore, for any integer \( k \in \{1, \dots, N+1\} \),
\[
    \mathbb{P}\left( s_{N+1}^{(j)} \leq s_{(k)}^{(j)} \right) = \frac{k}{N+1}.
\]

\noindent \textbf{Lower bound:} Setting \( k = \left\lceil (N+1)(1 - \alpha) \right\rceil \), we obtain:
\[
    \mathbb{P}\left( s_{N+1}^{(j)} \leq q_\alpha^{(j)} \right) = \frac{\left\lceil (N+1)(1 - \alpha) \right\rceil}{N+1}.
\]
Using the ceiling function property \( \lceil x \rceil \geq x \) for all \( x \in \mathbb{R} \), we have:
\[
    \frac{\left\lceil (N+1)(1 - \alpha) \right\rceil}{N+1} \geq \frac{(N+1)(1 - \alpha)}{N+1} = 1 - \alpha.
\]
Since the prediction set \( \mathcal{C}_\alpha^{(j)}(x_{N+1}) \) is defined such that:
\[
    y_{N+1}^{(j)} \in \mathcal{C}_\alpha^{(j)}(x_{N+1}) \iff s_{N+1}^{(j)} \leq q_\alpha^{(j)},
\]
it follows that:
\[
    \mathbb{P}\left( y_{N+1}^{(j)} \in \mathcal{C}_\alpha^{(j)}(x_{N+1}) \right) \geq 1 - \alpha.
\]

\noindent \textbf{Upper bound:} Since \( \left\lceil (N+1)(1 - \alpha) \right\rceil \leq (N+1)(1 - \alpha) + 1 \), we obtain:
\begin{align*}
    \mathbb{P}\left( y_{N+1}^{(j)} \in \mathcal{C}_\alpha^{(j)}(x_{N+1}) \right) 
    &= \mathbb{P}\left( s_{N+1}^{(j)} \leq q_\alpha^{(j)} \right) \\
    &= \frac{\left\lceil (N+1)(1 - \alpha) \right\rceil}{N+1} \\
    &\leq 1 - \alpha + \frac{1}{N+1}.
\end{align*}

\noindent \textbf{Conclusion:} Combining both bounds, we conclude that for each state dimension \( j \), the conformal prediction interval \( \mathcal{C}_\alpha^{(j)}(x_{N+1}) \) satisfies:
\[
    1 - \alpha \leq \mathbb{P}\left( y_{N+1}^{(j)} \in \mathcal{C}_\alpha^{(j)}(x_{N+1}) \right) \leq 1 - \alpha + \frac{1}{N+1}.
\]
This ensures marginal coverage validity per state dimension under standard split conformal assumptions.
\end{proof}

\begin{theorem}
    Under the same assumptions that the calibration scores \( \{s_{1}^{(j)}, s_{2}^{(j)}, \dots, s_{N}^{(j)} \} \) and the test score \( s_{N+1}^{(j)} \) are exchangeable and all scores are drawn from a continuous distribution, the marginal split conformal framework guarantees the following probabilistic error bounds for each state dimension \( j \in \{1, \dots, n\} \):
    \[
        1 - \alpha \leq \mathbb{P}\left( \left| y_{N+1}^{(j)} - \hat{y}_{N+1}^{(j)} \right| \leq \sqrt{ q_\alpha^{(j)} \cdot \hat{\sigma}_{N+1}^{(j)} } \right) \leq 1 - \alpha + \frac{1}{N+1}
    \]
\end{theorem}

\begin{proof}
\noindent \textbf{Conformity score and prediction interval:} For each state dimension \( j \), the conformity score for both calibration and test samples is defined as:
\[
    s_i^{(j)} = \frac{(y_i^{(j)} - \hat{y}_i^{(j)})^2}{\hat{\sigma}_i^{(j)}},
\]
and the corresponding conformal prediction interval is given by:
\[
    \mathcal{C}_\alpha^{(j)}(x_T) = \left[ \hat{y}_T^{(j)} - \sqrt{ q_\alpha^{(j)} \cdot \hat{\sigma}_T^{(j)} }, \;\; \hat{y}_T^{(j)} + \sqrt{ q_\alpha^{(j)} \cdot \hat{\sigma}_T^{(j)} } \right].
\]

\noindent \textbf{Equivalence of coverage and error events:} From the previous coverage proof, we know:
\[
    y_{N+1}^{(j)} \in \mathcal{C}_\alpha^{(j)}(x_{N+1}) \iff s_{N+1}^{(j)} \leq q_\alpha^{(j)}.
\]
Substituting the definition of the conformity score into this inequality gives:
\[
    \frac{(y_{N+1}^{(j)} - \hat{y}_{N+1}^{(j)})^2}{\hat{\sigma}_{N+1}^{(j)}} \leq q_\alpha^{(j)}.
\]
Multiplying both sides by \( \hat{\sigma}_{N+1}^{(j)} \) and taking square roots (since all terms are non-negative) yields:
\[
    \left| y_{N+1}^{(j)} - \hat{y}_{N+1}^{(j)} \right| \leq \sqrt{ q_\alpha^{(j)} \cdot \hat{\sigma}_{N+1}^{(j)} }.
\]
Therefore, the prediction error bound is equivalent to the coverage event:
\[
    y_{N+1}^{(j)} \in \mathcal{C}_\alpha^{(j)}(x_{N+1}) \iff \left| y_{N+1}^{(j)} - \hat{y}_{N+1}^{(j)} \right| \leq \sqrt{ q_\alpha^{(j)} \cdot \hat{\sigma}_{N+1}^{(j)} }.
\]

\noindent \textbf{Final error bound probabilistic guarantee:} By directly applying the result of the first theorem (i.e., the coverage guarantee for \( s_{N+1}^{(j)} \leq q_\alpha^{(j)} \)), we conclude:
\[
    1 - \alpha \leq \mathbb{P}\left( \left| y_{N+1}^{(j)} - \hat{y}_{N+1}^{(j)} \right| \leq \sqrt{ q_\alpha^{(j)} \cdot \hat{\sigma}_{N+1}^{(j)} } \right) \leq 1 - \alpha + \frac{1}{N+1}.
\]

\noindent \textbf{Conclusion:} This establishes a marginally valid and probabilistically guaranteed upper bound on the absolute prediction error in each dimension, scaled by the model's predicted standard deviation and the empirical quantile derived from calibration residuals.
\end{proof}
\subsubsection{Remarks on Marginal CP}
While our implementation ensures valid coverage and error bounds \emph{per dimension}, we acknowledge that the joint state vector $y_{N+1} \in \mathbb{R}^{n}$ may not be covered with probability $1 - \alpha$ due to the absence of joint CP construction such as the method found in \cite{romano2019conformalizedquantileregression}. However, marginal CP remains practical in robotics applications where the high probability coverage of individual states, especially when correlated, imply experimentally robust coverage of the entire joint vector $y_{N+1}$ despite theoretical loss \cite{Sampson2024Conformal}. In addition to this, to reflect the varying uncertainty that exists among different dynamic states (i.e. uncertainty in velocities versus uncertainty in accelerations), it remains our belief that computing an entire quantile vector $q_{\alpha} \in \mathbb{R}^{n}$ rather than scalar $q_{\alpha}$ being computed in the joint CP construction, to capture the statistical uncertainty of predicted states remains the more robust option. Nonetheless, the joint CP method from \cite{romano2019conformalizedquantileregression} will be included in the uncertainty quantification analyses within the Results \ref{sec:results}.

\section{Results and Validation}\label{sec:results}

In this section, we validate the PI-AttNP estimation approach by evaluating its performance on a simulated quadrotor perturbed by external forces, as well as additive and multiplicative noise from a multi-modal Gaussian distribution on the sensor measurements. We also verify that our approach augmented with CP-based uncertainty quantification enhances the distribution modeling capabilities of the PI-AttNP over the baselines we compared against in both training and inference through several evaluation metrics. 
%
\subsection{Baselines Used in Study} \label{subsec:baselines}
As mentioned prior, model baselines used to compare the performance of the PI-AttNP consist of the deep Kalman filter (DKF) and physics-informed neural network (PINN) approaches. Pertaining to the DKF, we adopt the model found in \cite{meng2024UKF} where we use transformers to model the transition and measurement models utilized in an inner UKF (unscented Kalman filter) algorithm to capture the nonlinear, possibly non-Gaussian behavior derived from dynamical model $f$. The uncertainty quantification for this model is derived from the statistical guarantees provided by the standard UKF augmented with learned covariance structure that varies at each time step. Note that we will also provide results from the standard UKF with hand-tuned covariance matrices as another baseline.

For the PINN, we employ the same dynamics model $g(\cdot)$ utilized by the PI-AttNP as a residual constraint in the PINN's optimization. From an architecture standpoint, we modify the PINN model discussed in \cite{Curto2024Pinn} where the PINN utilizes an LSTM with a Bayesian last layer (BLL) \cite{tran2019bayesian} at its head to enable more robust MC dropout to quantify the uncertainty of model predictions. This PINN-LSTM also employs self-attention between previous states $\hat{x}_k$ and control action $u_k$ to capture the interdependencies between these variables of interest to predict next states $\hat{x}_{k+1}$. 

In addition to these baselines, we also collect results from the standard attentive neural process (AttNP) to explicitly provide an ablation for physics model $g(\cdot)$ in the PI-AttNP's learning. To observe parameter counts and wall compute time for all models, please refer to Tables \ref{tab:param-counts} and \ref{tab:forward-pass} in the Appendix. For the direct model implementations, refer to our GitHub implementation link provided after Conclusions \ref{subsec:github-implementation}.

\subsection{Quadrotor Dynamics Implementation}
Let the general nonlinear vector field to be determined $f$, referenced in
\ref{eqn:dt-model} be the 6-DOF, underactuated dynamics of a quadrotor where $x \in \mathbb{R}^{12}$ is $x = \{\dot{r}^{I}, \dot{v}^{I}, \omega^{b}, \dot{\omega}^{b}\}$, $y_{k}\in\mathbb{R}^{6}$ is $y_{k}=\{\dot{v}_{k}^{I},\omega_{k}^{b}\}$, and $\xi \in \mathbb{R}^{6}$ represents a noise vector with unknown multimodal Gaussian noise characteristics possessing additive/multiplicative properties that perturb observed states $y_{k}$. Note that in our reformulation of $f$, we define translational velocities $\dot{r}^I \in \mathbb{R}^3$ and accelerations $\dot{v}^I \in \mathbb{R}^3$ in some defined inertial frame $I$. Additionally, we also define $f$ to possess the body rates of the quadrotor $\omega^b \in \mathbb{R}^3$ and its time derivative $\dot{\omega}^b \in \mathbb{R}^3$, where both of these quantities are expressed in the quadrotor's body frame $b$. Mathematically, these continuous-time dynamical states are modeled as:
\begin{equation}\label{eqn:model}
\begin{aligned}
    \dot{r}^I &= v^I \\
    \dot{v}^I &= \begin{bmatrix}
                    0 \\
                    0 \\
                    -g
                    \end{bmatrix} + \frac{1}{m} (R_{b}^{I} F_{m}^{b} + F_{\text{aero}}) = a^{I} \\
    \omega^{b} &= \begin{bmatrix}
                    1 & 0 & -s_{\theta} \\
                    0 & c_{\phi} & c_{\theta}s_{\phi} \\
                    0 & -s_{\phi} & c_{\theta}c_{\phi}
                    \end{bmatrix}  \dot{\Theta}^b  \\
    \dot{\omega}^b &= I^{-1}\left(\tau_{m} + \tau_{\text{aero}} - \tau_{g} - \omega^b \times I\omega^{b}\right),
\end{aligned}
\end{equation}
where $c_{\ast} = \text{cos}(\ast)$, $s_{\ast} = \text{sin}(\ast)$. We define $\dot{\Theta}^{b} \in \mathbb{R}^{3}$ as the time derivative of euler angles $\Theta^{b}=\{\phi,\theta,\psi\}\in\mathbb{R}^{3}$ where roll $\phi$, pitch $\theta$, and yaw $\psi$ are all expressed in body frame $b$. Additionally, $g$ is the acceleration due to gravity, $m$ is the mass of the quadrotor, $I \in \mathbb{R}^{3 \times 3}$ is the quadrotor's inertia tensor, $R_{b}^{I} \in \mathbb{R}^{3 \times 3}$ is a rotation matrix that transforms a vector in $b \rightarrow I$. It should be explicitly noted that for our purposes, we disregard position vector $r^{I}$, and we assume known, noisy Euler angles $\widetilde{\Theta}^{b}=\Theta^{b}+N(0,0.1)$ to enforce that all models are only being tasked to estimate the higher order dynamics as seen in equations \ref{eqn:model}.

Regarding modeled forces and torques, $F_{m}^{b} \in \mathbb{R}^3$ is the quadrotor's thrust vector expressed in the body frame that is produced from individual rotor velocities $\Bar{\omega} \in \mathbb{R}^4$, and $\tau_{m} \in \mathbb{R}^{3}$ are the torques that affect the quadrotor as a result of $\Bar{\omega}$. Along with these, $\tau_{g} \in \mathbb{R}^{3}$ are the gyroscopic torque effects. The internal forces and torques $F_{m}^{b},\;\tau_{m},\;\tau_{g}$ acting on the body are modeled as:
%
\begin{equation}
\begin{aligned} \label{eqn:modeled-forces}
    F_{m}^{b} &= k_{T} \sum_{i=1}^{4} \begin{bmatrix}
        0 \\
        0 \\
        \Bar{\omega}_{i}^{2}
    \end{bmatrix}, \quad \tau_{g} = \sum_{i=1}^{4} (-1)^{i+1} \Bar{\omega}_{i} \begin{bmatrix}
        I_{r} q I_{xx}^{-1}  \\
        I_{r} p I_{yy}^{-1} \\
        0
    \end{bmatrix}, \\[0.2cm]
    \text{and} \quad \tau_{m} &= \begin{bmatrix}
        \tau_{\phi} \\
        \tau_{\theta} \\
        \tau_{\psi}
    \end{bmatrix} = \begin{bmatrix}
        Lk_{T}(\Bar{\omega}_{1}^{2} - \Bar{\omega}_{3}^{2}) \\
        Lk_{T}(\Bar{\omega}_{2}^{2} - \Bar{\omega}_{4}^{2}) \\
        b\sum_{i=1}^{4}(-1)^{i+1}\Bar{\omega}_{i}^{2}
    \end{bmatrix},
\end{aligned}
\end{equation}
where $k_{T} > 0$ and $b > 0$ are the thrust and drag coefficients, respectively, $L > 0$ represents the distance between each rotor and the center of the quadrotor and $I_{r} \in \mathbb{R}$ represent the rotor moment of inertia. Also, $I_{xx}, I_{yy} \in \mathbb{R}$ are the first and second diagonal components of $I$, and $p,q \in \mathbb{R}$ are the $\phi$ and $\theta$ components of $\omega_{b}$.

$F_{\text{aero}}, \tau_{\text{aero}} \in \mathbb{R}^3$ observed in Equation \eqref{eqn:model} are the sum of external forces and torques acting on the quadrotor's body, respectively. Note that $F_{\text{aero}}$ and $\tau_{\text{aero}}$ are computed from $v_{w} \in \mathbb{R}^{3}$ (wind velocity) where components of $v_{w}$ are bounded such that $v_{w}\sim [v_{w(\text{min})}, v_{w(\text{max})}]$ and all of these quantities are modeled directly from the disturbance equations found in \cite{yu2023quadrotor}. Note that other nonlinear, external effects typically faced during quadrotor flight (i.e. blade flapping, induced drag, and ground-effect) are also modeled in $\Delta x_k$ according to the disturbance models described in \cite{yu2023quadrotor}. We also incorporate randomly-occuring rotor spiking phenomena where unobserved, uniformly-distributed rotor perturbations $\Bar{\omega} + \Delta\Bar{\omega}\in\mathbb{R}^{4}$ where $\Delta \Bar{\omega} \sim U(-|\Delta \Bar{\omega}_{\text{max}}|, |\Delta \Bar{\omega}_{\text{max}}|)$ are applied to observed rotor velocity signal $\Bar{\omega}$ given to models. In this way, we denote the main time-varying, perturbation parameters of the simulation $\Delta x_{k} = \{v_{w}, \Delta \Bar{\omega}_{k}\}$ along with modeled multimodal Gaussian noise $\xi_{k}$ applied to observed states $y_{k}$. To obtain a deeper understanding of the generation procedure for $\xi_k$ and how it is applied to observed states $y_k$, observe Algorithm \ref{alg:noise-implementation} and related discussions in the Appendix.

\begin{figure}
    \centering
    \setlength{\fboxsep}{0pt} 
    \setlength{\fboxrule}{0.5pt} 
    \fbox{\includegraphics[scale=0.4]{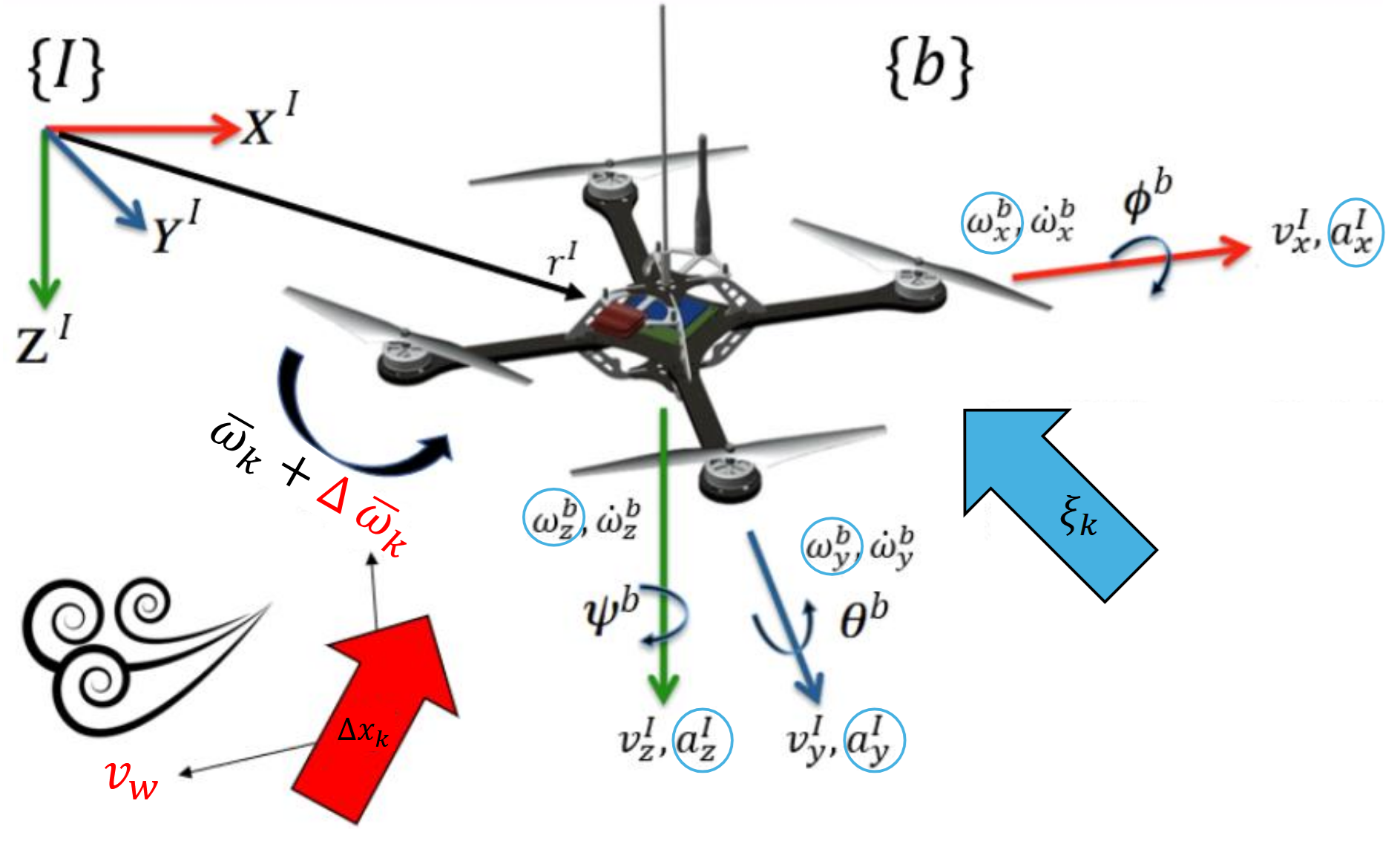}}
    \caption{Quadrotor dynamics and environment diagram where red denotes modeled external perturbations $\Delta x_{k}$ and blue denotes multimodal sensor noise $\xi_k$ introduced to observed states $y_k$}
    \label{fig:quadrotor-env-diagram}
\end{figure}

A representative diagram of the described simulated environment used in this study can be seen in Figure \ref{fig:quadrotor-env-diagram}. To observe the complete implementation of the quadrotor dynamics with incorporated noise vector $\xi_k$ and all state disturbances $\Delta x_k$, refer to our Github implementation with the link provided towards the end of the article \ref{subsec:github-implementation}. The perturbation limits for disturbances $\Delta x_k$ experienced during the generation of meta dataset $\mathcal{D}_{\text{meta}}$ can be observed in Table \ref{tab:perturb-bounds-new}. The main objective of our novel PI-AttNP algorithm estimates the states of the quadrotor system presented in Equations \eqref{eqn:model} and \eqref{eqn:modeled-forces} with no prior knowledge of sensor noise properties $\xi_k$ nor environmental disturbances acting on the quadrotor $\Delta x_k$. This indicates that the utilized kinematic model $g(\cdot)$ mentioned in the Problem Statement \ref{sec:problem-statement} and Algorithm \ref{alg:attnp-training-alg} consists of the equations found in \ref{eqn:model} and \ref{eqn:modeled-forces} with no knowledge of $F_{\text{aero}}$ nor $\tau_{\text{aero}}$.

\begin{table}[h!]
  \centering
  \begin{tabular}{|c|c|c|}
    \hline
     Sim Perturb Bounds & Min & Max \\
    \hline
    $v_w \;\; (m/s)$ & -30 & 30 \\
    \hline
    $\Delta \Bar{\omega}\;\;(rad/s)$ & -100 & 100 \\
    \hline
  \end{tabular}
  \caption{Perturbation bounds used for $\mathcal{D}_{\text{meta}}$}
  \label{tab:perturb-bounds-new}
\end{table}

\subsection{Context/Target Formulation}
In our discussion of the problem, note that the previously referenced context and target notation found in the Methodology \ref{sec:method} with $N_{C}=N_{T}=1$ describe the following:   
%
\begin{align*}
    x_{C} &= \left\{t_{k},\Bar{\omega}_{k} \right\} \in \mathbb{R}^{5} \\
    y_{C} &= \left\{\dot{r}_{k}^{I}, \dot{v}_{k}^{I}, \omega_{k}^{b}, \dot{\omega}_{k}^{b} \right\} \in \mathbb{R}^{12} \\
    x_{T} &= \left\{t_{k+1},\textbf{0} \right \} \in \mathbb{R}^{5}  \\
    y_{T} &= \left\{\dot{r}_{k+1}^{I}, \dot{v}_{k+1}^{I}, \omega_{k+1}^{b}, \dot{\omega}_{k+1}^{b} \right\} \in \mathbb{R}^{12},
\end{align*}
%
where $t_{k}$ and $t_{k+1}$ represents previous and next simulation time stamps and $\textbf{0}\in\mathbb{R}^{4}$ are padded zeros used to ensure size consistency between $x_{C}$ and $x_{T}$. The inclusion of $\Bar{\omega}_{k}$ in $x_{C}$ is placed there to represent the control action of the system $u_{k}$ due to the direct control effects that each rotor velocity has on the simulated quadrotor. Through this, the AttNP encodes all previous dynamic information $\phi_C$ and future time information in $x_{T}$ to predict desired dynamics $y_{T}$ at time step $k+1$. As previously mentioned in the recursive implementation section \ref{subsec:recursive-alg}, noise-perturbed observed states $y_{k}$ and posterior average $\hat{y}_{T}^{+}$ are fused to form $\widetilde{y}_{C}$ during trajectory estimation. Since we incorporate noisy observable data into the estimation, we adopt a similar prediction framework to other estimation algorithms where information from $\widetilde{y}_{C}$ is used to estimate latent dynamical states $y_{T}$ \cite{jin2021new,wang2017particle,Curto2024Pinn}.
%
%
\subsection{Error Metrics Used in Study} \label{subsec:metrics}
To objectively compute prediction error consistently with related literature \cite{Wan2000UKF,revach2022kalmannet,karl2017deepvariationalbayesfilters}, we use the root mean squared error (RMSE) between the 12 tracked quadrotor states. The formula for RMSE is given by:
\begin{align*}
    \text{RMSE} = \sqrt{\frac{1}{N} \sum_{i=1}^{N} \sum_{j=1}^{12} \left(Y_{i,j} - \hat{Y}_{i,j}\right)^{2}},
\end{align*}
where $N$ is the batch size, and $Y, \hat{Y} \in \mathbb{R}^{12}$ are the true and predicted quadrotor states, respectively.
In addition to the above metric, we utilize negative log-likelihood (NLL) as a metric to compare the capabilities of modeling marginal distribution $p(y_{T}|x_{T},\phi_{C})$ between the CP-augmented AttNP and previously-mentioned baselines. This metric is consistent with related literature in distribution modeling evaluation to ensure effects of distribution under/overconfidence are reflected \cite{kim2019attentiveneuralprocesses,revach2022kalmannet}. The formula for NLL is given by:
\begin{align*}
    \text{NLL} = \frac{1}{2N}\sum_{i=1}^{N}\sum_{j=1}^{12} \left(\log(2\pi \hat{\sigma}_{i,j}^{2}) + \frac{(Y_{i,j}-\hat{Y}_{i,j})^{2}}{\hat{\sigma}_{i,j}^{2}} \right)
\end{align*}

\subsection{Convergence Dynamics} \label{subsec:converge-dynamics}
\begin{figure*}[h]
    \centering
    
    \includegraphics[height=6cm,width=0.48\textwidth]{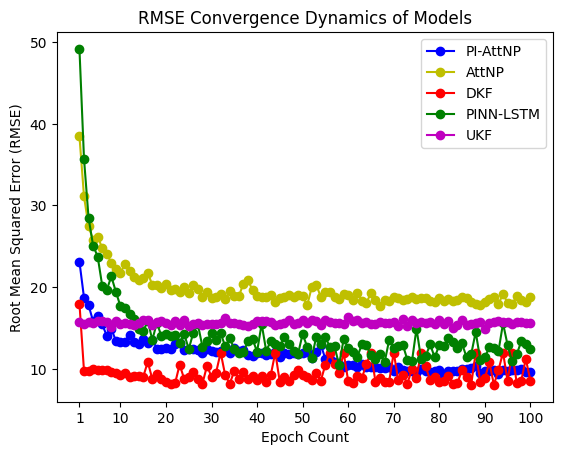}
    \hspace{0.01\textwidth}
    \includegraphics[height=6cm,width=0.48\textwidth]{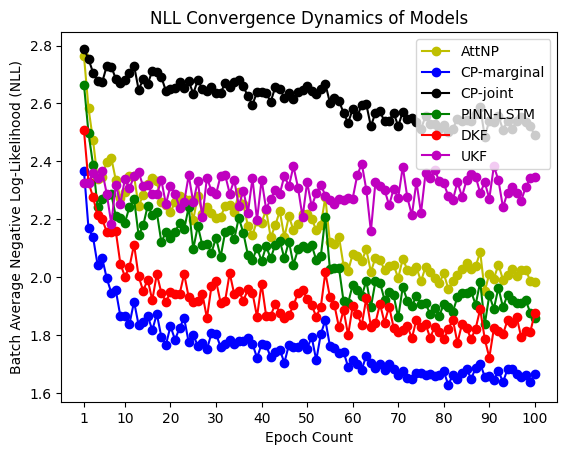}
    
    \caption{Model RMSE/NLL convergence dynamics of the PI-AttNP, AttNP, and additional baselines computed from randomly-sampled, unobserved minibatches $\mathcal{D}_{j}\sim\mathcal{D}_{\text{test}}$. Note that in the NLL dynamics plot, CP-marginal and CP-joint represent the results of the PI-AttNP wrapped with marginal and joint conformal prediction frameworks, respectively.}
    \label{fig:RMSE-NLL-results}
\end{figure*}
%
%
\begin{table}[h!]
  \centering
  \scriptsize  
  \renewcommand{\arraystretch}{1.3} 
  \begin{tabular}{|c|c|c|c|c|c|}
    \hline
     & PI-AttNP & AttNP & DKF & UKF & PINN-LSTM \\
    \hline
    Lowest RMSE & 9.845 $\pm$ 5.301 & 19.125 $\pm$ 8.654 & \textbf{8.657 $\pm$ 3.474} & 15.105 $\pm$ 6.604 & 11.484 $\pm$ 7.421 \\
    \hline
  \end{tabular}
  \caption{Lowest RMSE reached for all models in the plot to the left within Figure \ref{fig:RMSE-NLL-results}}
  \label{tab:train-RMSE-results}
\end{table}

\begin{table}[h!]
  \centering
  \renewcommand{\arraystretch}{1.3} 
  {\fontsize{7}{8}\selectfont
  \begin{tabular}{|c|c|c|c|c|c|c|}
    \hline
     & CP-Marginal & CP-Joint & AttNP & DKF & UKF & PINN-LSTM \\
    \hline
    {\fontsize{9}{10}\selectfont Lowest NLL} & \textbf{1.615 $\pm$ 0.251}
               & 2.523 $\pm$ 0.118
               & 1.958 $\pm$ 1.504
               & 1.757 $\pm$ 0.602
               & 2.173 $\pm$ 1.536
               & 1.896 $\pm$ 1.472 \\
    \hline
  \end{tabular}
  }
  \caption{Lowest NLL reached for all models in the plot to the right within Figure \ref{fig:RMSE-NLL-results}}
  \label{tab:train-NLL-results}
\end{table}

\begin{figure}[h]
    \centering
    \begin{minipage}[t]{0.48\textwidth}
        \centering
        \includegraphics[height=6cm,width=\linewidth]{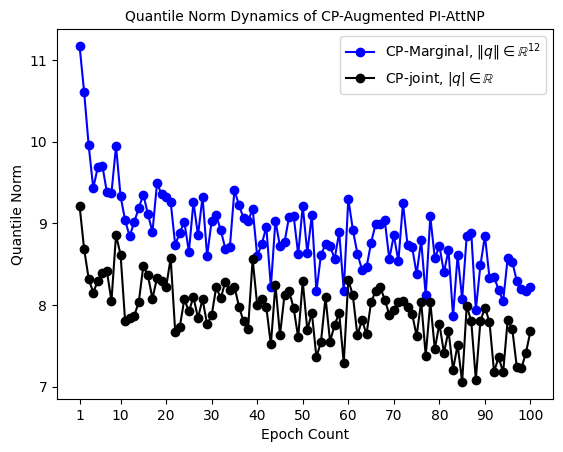}
    \end{minipage}%
    \hfill
    \begin{minipage}[t]{0.48\textwidth}
        \centering
        \includegraphics[height=6cm,width=\linewidth]{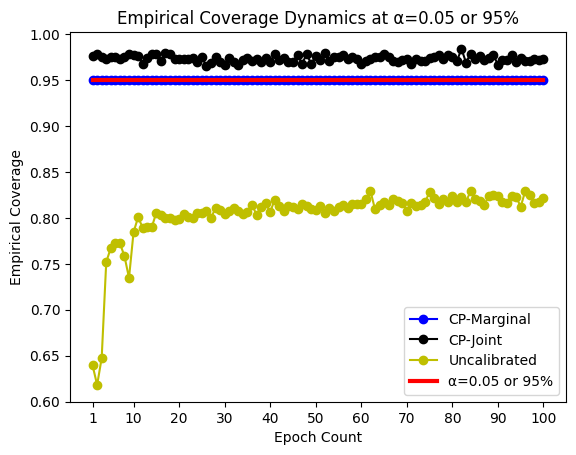}
    \end{minipage}
    
    \caption{Quantile norm and empirical coverage dynamics plots for the marginal CP-augmented PI-AttNP and joint CP-augmented PI-AttNP approaches. Note that for the coverage plot on the right, quantiles for model uncertainty $q_{\alpha}$ were computed based on $\alpha=0.05$ and $|\mathcal{D}_{\text{cal}}|=1000$. Given samples $(x_{i},y_{i})\sim \mathcal{D}_{\text{cal}}$, we denote coverage $=\;(|y_{i}\in\mathcal{C}(x_{i})| \;/\;|\mathcal{D}_{\text{cal}}|) \times 100\%$. For CP-marginal, the coverage computed for each individual state was averaged together to get the scalar value observed in the plot.}
    \label{fig:q-norm-traj}
\end{figure}

\begin{table}[h!]
  \centering
  \small
  \begin{tabular}{|c|c|c|c|}
    \hline
     & CP-Marginal & CP-Joint & Uncalibrated \\
    \hline
    Mean Coverage & 0.9499 $\pm$ 0.0000 & 0.9730 $\pm$ 0.0038 & 0.8017 $\pm$ 0.0332 \\
    \hline
  \end{tabular}
  \caption{Average experimental coverage through training for CP variations of the PI-AttNP}
  \label{tab:mean-coverage}
\end{table}


\begin{figure*}[h]
    \centering
    
    \includegraphics[height=5cm,width=0.45\textwidth]{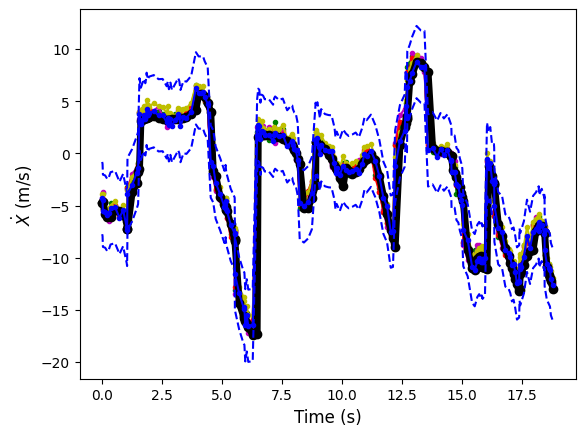}
    \hspace{0.01\textwidth}
    \includegraphics[height=5cm,width=0.45\textwidth]{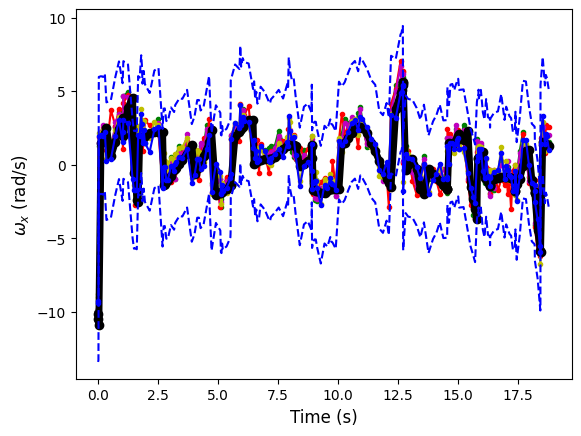}
    \vspace{0.7em}
    
    \hspace{0.01\textwidth}
    \includegraphics[height=5cm,width=0.45\textwidth]{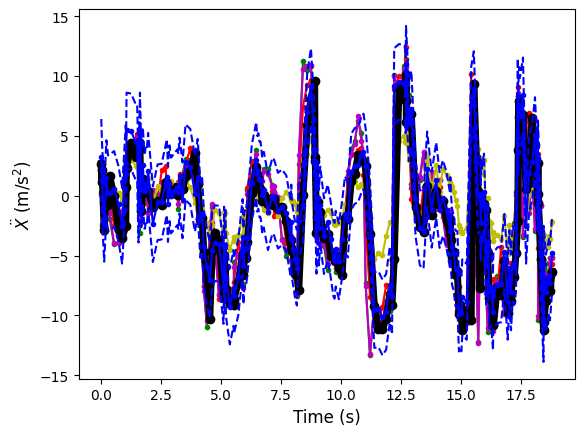}
    \hspace{0.01\textwidth}
    \includegraphics[height=5cm,width=0.45\textwidth]{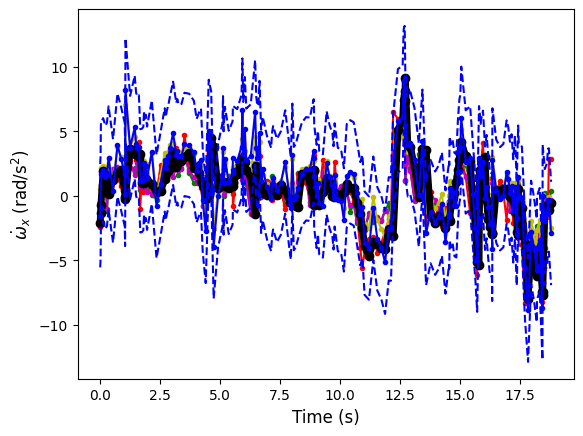}
    
    \caption{X-direction translational velocity and acceleration $v_{x}^{I},a_{x}^{I}$ (first column) along with X-direction rotational velocity and acceleration $\omega_x, \dot{\omega}_x$ state trajectory comparisons to ground-truth (black) between the PI-AttNP (blue) w/ marginal CP-based confidence bounds $\pm q_{\alpha}\cdot\hat{\sigma}_{T}$ (dotted-blue), AttNP (yellow) \cite{kim2019attentiveneuralprocesses}, Deep Kalman Filter (red) \cite{meng2024UKF}, PINN-LSTM (green) \cite{Curto2024Pinn}, and hand-tuned covariance UKF (magenta). Note that perturbation parameters used above are $v_w = 50 \begin{bmatrix}
        1 & -1  &1
    \end{bmatrix} \;m/s$, $|\Delta \Bar{\omega}_{\text{max}}| = 200\;rad/s$. Observe numerical results of all models in Table \ref{tab:traj-RMSE-results}, and observe the remaining state trajectory plots in the Y/Z directions in Figure \ref{fig:remain-traj-plots}.}
    \label{fig:initial-traj-plots}
\end{figure*}

\newcommand{\numsmaller}[1]{{\fontsize{9}{10}\selectfont #1}} 

\begin{table}[h]
  \centering
  \small 
  \begin{tabular}{|c|c|c|c|c|}
    \hline
    {\scriptsize Model RMSE} & $\dot{r}^{I}$ & $\dot{v}^{I}$ & $\omega^{b}$ & $\dot{\omega}^{b}$ \\
    \hline
    {\scriptsize PI-AttNP} & \textbf{\numsmaller{1.215 $\pm$ 0.593}} & \numsmaller{2.784 $\pm$ 1.550} & \textbf{\numsmaller{0.481 $\pm$ 0.255}} & \numsmaller{4.211 $\pm$ 2.354} \\
    \hline
    {\scriptsize AttNP} & \numsmaller{1.356 $\pm$ 0.690} & \numsmaller{8.726 $\pm$ 3.913} & \numsmaller{1.887 $\pm$ 0.955} & \numsmaller{4.865 $\pm$ 2.844} \\
    \hline
    {\scriptsize DKF} & \numsmaller{1.428 $\pm$ 0.678} & \textbf{\numsmaller{1.787 $\pm$ 0.805}} & \numsmaller{1.336 $\pm$ 0.771} & \textbf{\numsmaller{2.565 $\pm$ 1.637}} \\
    \hline
    {\scriptsize LSTM-PINN} & \numsmaller{1.649 $\pm$ 0.574} & \numsmaller{4.527 $\pm$ 2.104} & \numsmaller{1.576 $\pm$ 0.583} & \numsmaller{3.929 $\pm$ 2.013} \\
    \hline
    {\scriptsize UKF} & \numsmaller{1.854 $\pm$ 0.701} & \numsmaller{4.015 $\pm$ 2.721} & \numsmaller{1.707 $\pm$ 0.243} & \numsmaller{6.590 $\pm$ 3.280} \\
    \hline
  \end{tabular}
  \caption{Estimation RMSE Results from Models in Figures \ref{fig:initial-traj-plots} and \ref{fig:remain-traj-plots}. Note that all RMSE values computed for vector quantities $\dot{r}^{I}, \dot{v}^{I}, \omega^{b},\dot{\omega}^{b}$, we use the RMSE formula found in \ref{subsec:metrics} with $N=3$ for the X-Y-Z components of each dynamic state.}
  \label{tab:traj-RMSE-results}
\end{table}
From our analyses, it can be shown that the PI-AttNP possesses the capability of learning quadrotor dynamics at a substantial level. This is shown explicitly in the observation of the graphs provided in the convergence dynamics in Section \ref{subsec:converge-dynamics}, and in the test trajectory results in Table \ref{tab:traj-RMSE-results}, where the quality of state predictions from the PI-AttNP rivals  the DKF with respect to the lowest RMSE error and training convergence by notable margins compared to the original AttNP and other baselines. See, for example, the left image in Figure \ref{fig:RMSE-NLL-results}. We also note that when observing the NLL results in the plot to the right in Figure \ref{fig:RMSE-NLL-results}, our proposed marginal CP-based PI-AttNP provides the best results in the distributional modeling of desired predictive distribution $p(y_{T}|x_{T},\phi_C)$ in comparsion to all other baselines with the DKF coming closely in 2nd-place. It should be explicitly noted, however, that when observing the model capacity of both the PI-AttNP and DKF in Table \ref{tab:param-counts}, the PI-AttNP is much smaller than the DKF by more than $15\times$ the magnitude - indicating more feasibility in real-time, low-compute applications. Our comparison with the hand-tuned UKF shows the difficulty of manually capturing valid process and measurement covariance matrices in complex, uncertain environments - an issue that is mitigated by the PI-AttNP due to its data-driven nature. 

When observing the quantile norm trajectories through model training shown in the image to the left within Figure \ref{fig:q-norm-traj}, we can observe that the quantile magnitudes adaptively decrease as training progresses for both the marginal and joint CP approaches. This is expected as the model improves, the computed nonconformity scores found in Equation \eqref{eq:conformity-score} should become smaller. From our coverage analysis shown in the image to the right within Figure \ref{fig:q-norm-traj}, it should be understood that the empirical coverage is computed as the relative frequency for the number of samples in a randomly-sampled calibration set (i.e. $\mathcal{D}_{\text{cal}}\sim\mathcal{D}_{\text{test}}$) where the ground-truth $y_{T}$ is contained within one predicted standard deviation of the predicted mean (i.e. for samples in $\mathcal{D}_{\text{cal}}$, how many $y_{T}\in\hat{y}_{T} \pm q_{\alpha}\cdot \hat{\sigma}_{T}$ ?). From this analysis and the results shown in Table \ref{tab:mean-coverage}, the marginal conformal prediction algorithm provides superior coverage over both the joint CP and uncalibrated (i.e. $q_{\alpha}=1$) variations of the PI-AttNP for providing calibrated uncertainty bounds with statistical guarantees based on our user-defined coverage parameter $\alpha=0.05$, thus supporting exact theoretical coverage guarantees provided by Theorem \ref{thm:coverage}. It should be especially noted that the near perfect coverage provided by marginal CP was achieved even during early epochs - indicating fast adaptability for capturing quantified uncertainty in model predictions. These analyzed qualities from marginal CP used in tandem with our novel PI-AttNP approach provides further evidence that our algorithm can be applied in real-time, active learning settings where initial estimates may be inaccurate, but we can still obtain quantified uncertainty. 

Through these experimental analyses, the PI-AttNP has shown indication that it can learn the dynamics for a system subject to significant external disturbances and still provide accurate estimation with robust guarantees provided from the CP framework. We also observe from our experimentation that our novel inclusion of a simplified kinematics model $g(\cdot)$ clearly provides added benefit over the previous vanilla AttNP. 

\subsection{Remarks on Data Exchangability in Estimation}

\begin{figure*}[h]
    \centering
    
    \includegraphics[height=5.5cm,width=0.48\textwidth]{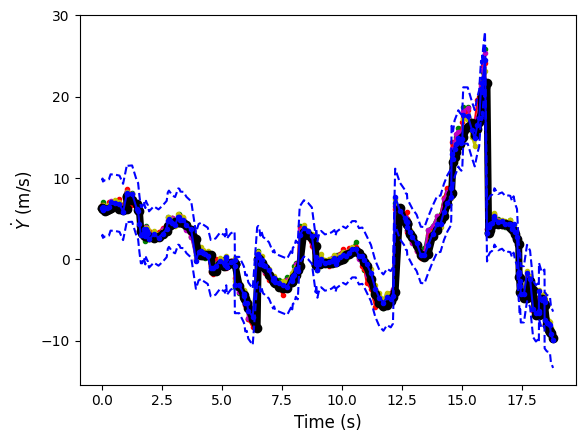}
    \hspace{0.01\textwidth}
    \includegraphics[height=5.5cm,width=0.48\textwidth]{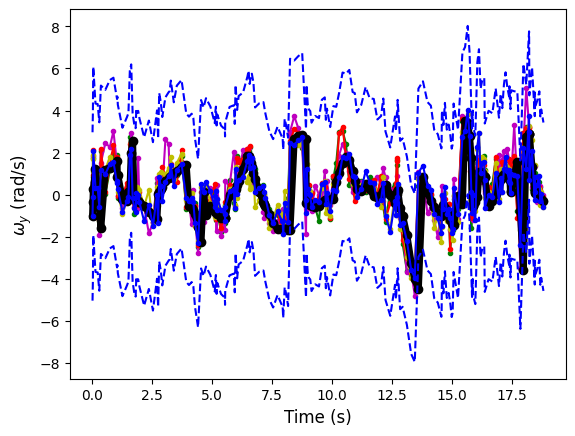}
    \hspace{0.01\textwidth}
    \includegraphics[height=5.5cm,width=0.48\textwidth]{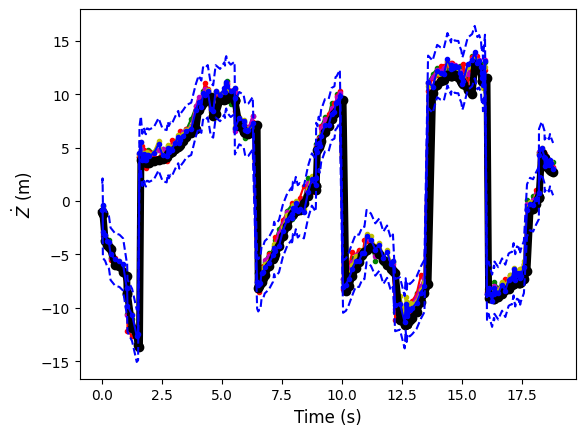}
    \hspace{0.01\textwidth}
    \includegraphics[height=5.5cm,width=0.48\textwidth]{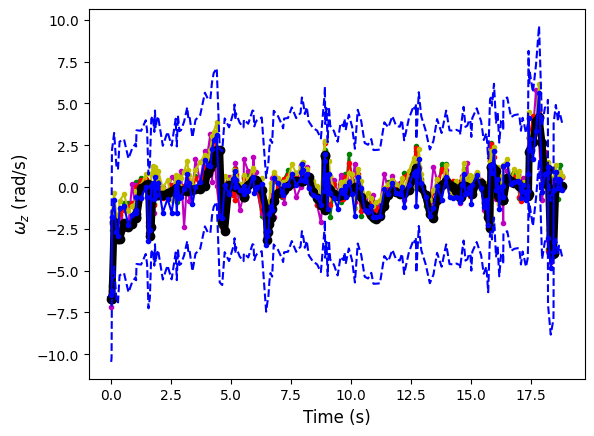}
    
    \vspace{0.7em}
    
    \includegraphics[height=5.5cm,width=0.48\textwidth]{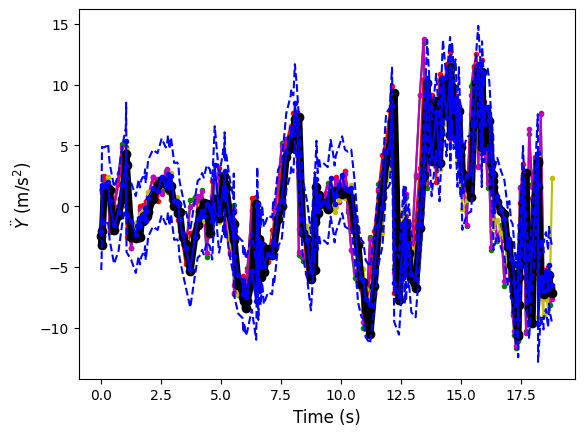}
    \hspace{0.01\textwidth}
    \includegraphics[height=5.5cm,width=0.48\textwidth]{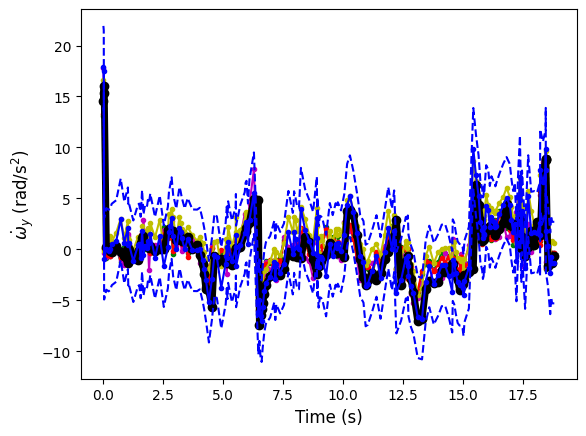}
    \hspace{0.01\textwidth}
    \includegraphics[height=5.5cm,width=0.48\textwidth]{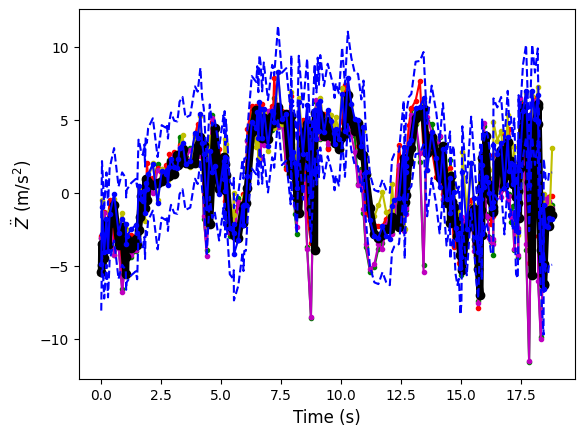}
    \hspace{0.01\textwidth}
    \includegraphics[height=5.5cm,width=0.48\textwidth]{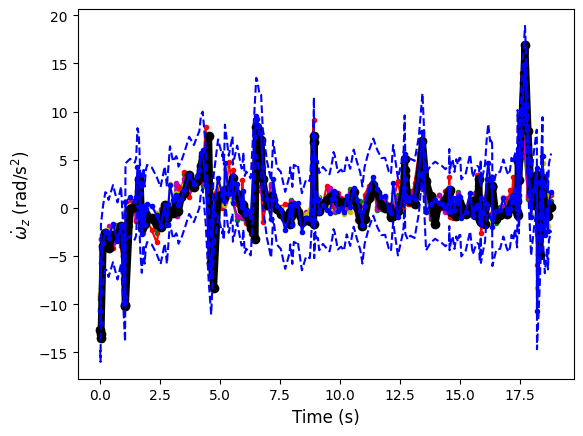}
    
    \caption{Y/Z-direction translational velocities and accelerations $v_{y}^{I},v_{z}^{I},a_{y}^{I},a_{z}^{I}$ along with Y/Z-direction rotational velocity and accelerations $\omega_{y}^{b},\omega_{z}^{b},\dot{\omega}_{y}^{b},\dot{\omega}_{z}^{b}$ state trajectory comparisons to ground-truth (black) between the PI-AttNP (blue) w/ CP-based confidence bounds $\pm q_{\alpha}\cdot\sigma_{T}$ (dotted-blue), AttNP (yellow) \cite{kim2019attentiveneuralprocesses}, Deep Kalman Filter (red) \cite{meng2024UKF}, PINN-LSTM (green) \cite{Curto2024Pinn}, and hand-tuned covariance UKF (magenta). Note that perturbation parameters used above are $v_w = 50 \begin{bmatrix}
        1 & -1  &1
    \end{bmatrix} \;m/s$, $|\Delta \Bar{\omega}_{\text{max}}| = 200\;rad/s$. Observe numerical results of all models in Table \ref{tab:traj-RMSE-results}.}
    \label{fig:remain-traj-plots}
\end{figure*}

As mentioned previously in the CP section \ref{subsec:cp}, a key assumption of conformal prediction is that test sample $s_{N+1}$ remains exchangable with calibration scores from set $\mathcal{S}=\{s_{1},s_{2},\dots,s_{N}\}$ \cite{Sampson2024Conformal}. This entails that the data distribution by which $s_{N+1}$ is generated from is consistent with the distribution that $\mathcal{S}$ were all generated from. Due to this, only the latest sensor readings must be utilized as data within $\mathcal{D}_{\text{cal}}$ to ensure only recent sensor noise behavior is reflected in the computation of $q_{\alpha}$. When observing the predicted $\pm q_{\alpha} \cdot \sigma_{T}$ bounds found in the plots within Figures \ref{fig:initial-traj-plots} and \ref{fig:remain-traj-plots}, it is noted that the test state trajectories were generated from a different underlying data distribution that $\mathcal{D}_{\text{train}}$ was generated from due to all models being trained using data with lower levels of disturbance applied. Therefore, the exchangability requirement is potentially violated between any incoming, arbitrary test point $(x_{T},y_{T})$ and $\mathcal{D}_{\text{cal}}$. Despite this, we still observe tight, generally accurate quantile-scaled bounds coming from our PI-AttNP that remain viable in an estimation task. While this is the case for this study, we acknowledge that the failure of the exchangability assumption in out of distribution (OOD) cases can cause for misrepresented uncertainty. Practically, this may happen when sensor noise behavior recorded from sample $s_{N+1}$ may differ from $\mathcal{S}$ due to abrupt noise or disturbance changes in the model's environment as could occur in real-world settings. Based on this, an \emph{adaptive} conformal prediction framework must be adopted to ensure theoretical probabilistic coverage and error bounds remains despite adverse changes to the environment.  
\section{Conclusion and Future Works}\label{sec:conclude+future}
Through observation of the PI-AttNP's estimation results within the provided analyses compared to other model baselines, it has been shown that the PI-AttNP provides competitive performance in comparison to the state-of-the-art. While additional prevalent noise characteristics (i.e. bias drift, exponential, etc.) can be applied to verify our claim further, these results have provided promise to our methodology. While our current validation was limited to a high-fidelity simulation in this study, we view these results as an essential precursor to physical deployment. As a future research direction, we wish to implement the PI-AttNP estimation approach within a physical robot to truly validate our proposed algorithm for real-time applications. We also wish to incorporate a learned measurement model $\hat{h}(\cdot)$ to provide more robust error correction for predicted latent states. Additionally, as a proposed real-time approach, our developed CP-based uncertainty quantification method must be leveraged \emph{adaptively} throughout a mission, especially for safety-critical applications where we require the estimation error to be bounded. Finally, we desire to minimize the discrepancy between low-fidelity physics model $g(\cdot)$ and ground-truth dynamics observed from $y_{T}$ by approximating the \emph{dynamic residuals} such as in the approach found in \cite{niu2024multifidelityresidualneuralprocesses}. This novel, real-time, safety-critical PI-AttNP model formulation could be extended to a wider range of applications such as multi-agent estimation, model-based reinforcement learning, and model predictive control.

\subsection*{Acknowledgements}
The authors would like to thank the Florida Educational Fund and the McKnight Foundation for supporting this work.

\subsection*{Funding}
This research was funded, in part, by the Florida Educational Fund and the McKnight Foundation.

\subsection*{Data Availability} \label{subsec:github-implementation}
The datasets generated and the implementation of the proposed PI-AttNP estimator, together with the simulation quadrotor environment used in this study, are openly available in the GitHub repository at \url{https://github.com/devin1126/AI-Driven-State-Estimation-for-Simulated-Quadrotors}.

\subsection*{Conflict of Interest}
The authors declare that they have no conflict of interest.

\subsection*{Author Contributions}
Devin Hunter: Conceptualization, Methodology, Software, Validation, Writing – Original Draft, Visualization.  
Chinwendu Enyioha: Supervision, Methodology, Writing – Review $\&$ Editing, Resources.  
Both authors contributed equally to this work.

\section{Appendix} \label{sec:appendix}
\subsection{Marginal \& Joint CP Implementations}
In observing the algorithms found in \ref{alg:marginal-cp-alg} and \ref{alg:joint-cp-alg}, we demonstrate the computation of user-defined $(1-\alpha)$ quantiles used to construct  conformal set $\mathcal{C}_{\alpha}(x_{T})$ in both the marginal and joint versions of this algorithm. We refer readers to Section \ref{subsec:cp} for more details on the conformal prediction algorithm. For both approaches, note that we always used a calibration set size $|\mathcal{D}_{\text{cal}}|=1000$ after each epoch to ensure our computed quantile approximated the model's true uncertainty to an adequate level. Pertaining to the marginal algorithm, it is important to note that in specifically computing $q_{\alpha}^{(j)}$, the quantity $\mathcal{S}^{(j)}[\left\lceil (N+1)(1-\alpha) \right\rceil]$ seen in line 17 represents the $\left\lceil (N+1)(1-\alpha) \right\rceil$-th entry of the state-specific ordered array $\mathcal{S}^{(j)}$. Additionally, recall that $n=\text{state-dim}(y_{T})$. Therefore, it is clear that in the joint version of the CP algorithm, we note that the conformity score computation in line 13 uses the entire state vector for both $y_{T}$ and $\hat{y}_{T}^{+}$ to evaluate scalar quantile $q_\alpha$. In observing the conformity score computation, we note that $\left(\sqrt{\hat{\sigma}_{T}^{2}} \cdot I\right)^{-1}\in\mathbb{R}^{n\times n}$ represents an inverted identity matrix where each diagonal term is scaled by each model predicted standard deviation in $\hat{\sigma}_{T}\in\mathbb{R}^{n}$.  

\subsection{Real-Time Feasibility Discussion}

When observing the parameter counts of all models in Table \ref{tab:param-counts}, it can be seen that the AttNP contains significantly less parameters in comparison to the baselines despite being a purely data-driven approach (note that physics model $g(\cdot)$ does not increase parameter count of AttNP). It should also be noted that the PI-AttNP and other baselines all possess parameter counts low enough to generally be utilized in many forms of micro-processing hardware such as a Raspberry Pi, NVIDIA Jetson, etc \cite{baller2021deepedgebenchbenchmarkingdeepneural}. 

To further emphasize the feasibility of our models in a real-time setting, we record the compute time for each model's forward pass of $N = 1,100, \;\text{and}\;1000$ simultaneous samples, respectively. This simulation of parallel forward passes can be seen as a representative example of the compute time needed in multi-agent state estimation. As can be gathered from Table \ref{tab:forward-pass}, our PI-AttNP approach remains an attractive option in comparison to other baselines in computational efficiency. We observe that the PINN-LSTM generally provides the fastest single foward pass inference time across all batch sizes; however, with the incorporation of MC dropout, computation time will scale linearly by the number of forward passes used in computing the model's predictive posterior - a computational bottleneck that is not required for the PI-AttNP. Pertaining to the DKF, we observe that due to the size of the model and recursive UKF algorithm utilized, computation times are considerably larger than other approaches. Note that all recorded times were executed on a computer with an 13th Gen Intel® Core™ i7-1360P × 16 with 32 GB of RAM.  

\begin{table}[h!]
  \centering
  \large
  \begin{tabular}{|c|c|}
    \hline
    Models & Parameter Count \\
    \hline
    PI-/AttNP & 192408 \\
    \hline
    LSTM-PINN & 446732 \\
    \hline
    DKF & 3183384 \\
    \hline
  \end{tabular}
  \caption{Parameter Counts for Each Model}
  \label{tab:param-counts}
\end{table}

\begin{table}[h!]
  \centering
  \scalebox{1.2}{ 
    \begin{tabular}{|c|c|c|c|}
      \hline
      Models & Time (s) & Time (s) & Time (s) \\
      \hline
      PI-AttNP & 0.0040 $\pm$ 0.0011 & 0.0116 $\pm$ 0.0022 & 0.0356 $\pm$ 0.0029 \\
      \hline
      AttNP & 0.0034 $\pm$ 0.0016 & 0.0063 $\pm$ 0.0027 & 0.0208 $\pm$ 0.0173 \\
      \hline
      DKF & 0.0172 $\pm$ 0.0118 & 0.2683 $\pm$ 0.0479 & 2.4177 $\pm$ 0.0804 \\
      \hline
      LSTM-PINN & 0.0011 $\pm$ 0.0013 & 0.0069 $\pm$ 0.0125 & 0.0280 $\pm$ 0.0134 \\
      \hline
      Sample Size & $N=1$ & $N=100$ & $N=1000$ \\
      \hline
    \end{tabular}%
  }
  \caption{Runtime for $N = 1,100,\;\text{and}\;1000$ Parallel Forward Passes}
  \label{tab:forward-pass}
\end{table}

\subsection{Noise Generation Implementation for $\xi_k$}

In Algorithm \ref{alg:noise-implementation}, one can determine how multimodal Gaussian noise vector $\xi_k$ is computed through a sampling mechanism and injected into $y_{k}$ with both additive/multiplicative components similar to other approaches \cite{imuZheng2023}. From the input arguments, we denote $n_{\text{max}}$ the maximum number of peaks that can exist in $\xi_k$, $N_{\text{total}}$ as the total number of sample points stored in arbitrary array $\mathcal{G}$, and $s_{\text{motion}}$ being a small scalar used to ensure the incorporation of multiplicative effects from $y_k$ remain within realistic bounds for typical motion sensors. Additionally, note that Gaussian noise parameter minimum/maximum values $\mu_{\xi(\text{min})},\; \mu_{\xi(\text{max})},\sigma_{\xi(\text{min})},\; \sigma_{\xi(\text{max})}$ are utilized as bounds by which observed state Gaussian parameters $\mu_{\text{peak}},\sigma_{\text{peak}}$ are uniformly sampled. These parameters were sampled to model the variation in sensor performance that could occur in the utilization of multiple sensors with varying noise characteristics \cite{noiseWang2019}. In all previous analyses, we explicitly set $n_{\text{max}}=5$, $N_{\text{total}}=100$, and $s_{\text{motion}}=10^{-4}$. Explicitly-defined bounds for $\xi_k$ (i.e. $\mu_{\xi(\text{min})},\; \mu_{\xi(\text{max})},\sigma_{\xi(\text{min})},\; \sigma_{\xi(\text{max})}$) are summarized in Table \ref{tab:trans-ang-sensor-noise-params}. The parameter values selected represent extremely noisy IMU (inertial measurement unit) sensors and are utilized primarily as a stress test to show how robust all approaches are in uncertain sensing regimes \cite{imuZheng2023,park2008imu}.

\begin{table}[h!] 
  \centering
  \begin{tabular}{|c|c|c|c|c|}
    \hline
    Bounds for $\xi$ & $\mu_{a}$ & $\sigma_{a}$ & $\mu_{\omega}$ & $\sigma_{\omega}$\\
    \hline
    Max Values & 0.75 & 0.9 & 0.5 & 1.0 \\
    \hline
    Min Values & 0.075 & 0.015 & 0.01 & 0.05  \\
    \hline
    Units & $m/s^{2}$ & $m/s^{2}$ & $rad/s$ & $rad/s$ \\
    \hline
  \end{tabular}
  \caption{Simulated Translational \& Angular Noise Bounds} 
  \label{tab:trans-ang-sensor-noise-params}
\end{table}

\begin{algorithm}
\caption{Marginal Conformal Prediction Algorithm}
\label{alg:marginal-cp-alg}
\begin{algorithmic}[1]
\Require $\alpha,\;\mathcal{D}_{\text{cal}},\; \hat{f}_{\Gamma}(\cdot),\;g(\cdot)$
\Ensure $\mathcal{D}_{\text{cal}} \subsetneq\;  \mathcal{D}_{\text{train}}$
\For{$j$ \textbf{in} $n$}
\For{$i$ \textbf{in} $\mathcal{D}_{\text{cal}}$}
\State {\small \textbf{Parsing Context and Target Sets from Current $\mathcal{D}_{i}$}}
\State $x_{C},\widetilde{y}_{C}^{(j)}, x_{T}, y_{T}^{(j)} \leftarrow \mathcal{D}_{i}$ \\
\State $\textbf{Computing Apriori Estimate over}$
\State $\textbf{Next States}$
\State $\hat{y}_{T}^{-(j)} \gets g(x_{C},\widetilde{y}_{C}^{(j)},x_{T})$ \\
\State $\textbf{Computing Predictive Distribution over}\; y_{T}$
\State $\hat{y}_{T}^{+(j)},\hat{\sigma}_{T}^{2(j)} \gets \hat{f}_{\Gamma}(x_{C},\widetilde{y}_{C}^{(j)},x_{T},\hat{y}_{T}^{-(j)})$ \\
\State $\textbf{Compute Conformity Score of} \;i \textbf{-th}\; \textbf{Sample}\;$
\State $s_{i}^{(j)} \gets \left(y_{T}^{(j)} - \hat{y}_{T}^{+(j)}\right)^{2}\;/\; {\sqrt{\hat{\sigma}_{T}^{2(j)}}}$ \\
\State $\textbf{Store Score Sample in Ordered Set }\mathcal{S}^{(j)}$
\State $\mathcal{S}^{(j)}  \gets s_{i}^{(j)}$
\State $(\textbf{Note that:}\;\; \mathcal{S}^{(j)} = \{s_{0}^{(j)},s_{1}^{(j)},\dots,s_{i-1}^{(j)}\})$ 
\EndFor
\State $\textbf{Compute}\;\;(1-\alpha)\;\; \textbf{State Coverage Quantile}$
\State $q_{\alpha}^{(j)}\gets \mathcal{S}^{(j)}[\left\lceil (N+1)(1-\alpha) \right\rceil]$
\EndFor
\State $\textbf{return}\;\;q_{\alpha}\in\mathbb{R}^{n}$
\end{algorithmic}
\end{algorithm}

\begin{algorithm}
\caption{Joint Conformal Prediction Algorithm}
\label{alg:joint-cp-alg}
\begin{algorithmic}[1]
\Require $\alpha,\;\mathcal{D}_{\text{cal}},\; \hat{f}_{\Gamma}(\cdot),\;g(\cdot)$
\Ensure $\mathcal{D}_{\text{cal}} \subsetneq\;  \mathcal{D}_{\text{train}}$
\For{$i$ \textbf{in} $\mathcal{D}_{\text{cal}}$}
\State $\textbf{Parsing Context and Target Sets from Current $\mathcal{D}_{i}$}$
\State $x_{C},\widetilde{y}_{C}, x_{T}, y_{T} \leftarrow \mathcal{D}_{i}$ \\
\State $\textbf{Computing Apriori Estimate over}$
\State $\textbf{Next States}$
\State $\hat{y}_{T}^{-} \gets g(x_{C},\widetilde{y}_{C},x_{T})$ \\
\State $\textbf{Computing Predictive Distribution over}\; y_{T}$
\State $\hat{y}_{T}^{+},\hat{\sigma}_{T}^{2} \gets \hat{f}_{\Gamma}(x_{C},\widetilde{y}_{C},x_{T},\hat{y}_{T}^{-})$ \\
\State $\textbf{Compute Conformity Score of} \;j \textbf{-th}\; \textbf{Sample}\;$
\State $s_{i} \gets \left(y_{T} - \hat{y}_{T}^{+}\right)^{T} \left({\sqrt{\hat{\sigma}_{T}^{2}}\;\cdot I} \right)^{-1}\left(y_{T} - \hat{y}_{T}^{+}\right)$ \\
\State $\textbf{Store Score Sample in Ordered Set }\mathcal{S}$
\State $\mathcal{S}  \gets s_{i}$
\State $(\textbf{Note that:}\;\; \mathcal{S} = \{s_{0},s_{1},\dots,s_{i-1}\})$ 
\EndFor
\State $\textbf{Compute}\;\;(1-\alpha)\;\; \textbf{State Coverage Quantile}$
\State $q_{\alpha}\gets \mathcal{S}[\left\lceil (N+1)(1-\alpha) \right\rceil]$
\State $\textbf{return}\;\;q_{\alpha}\in\mathbb{R}$
\end{algorithmic}
\end{algorithm}

\begin{algorithm}[!h]
\caption{Noise Generation Algorithm for $\xi_{k}$ and Injection into $y_{k}$}
\label{alg:noise-implementation}
\begin{algorithmic}[1]
\Require $y_{k},\;n_{\text{max}},\;N_{\text{total}},\;\mu_{\xi(\text{min})},\; \mu_{\xi(\text{max})},\sigma_{\xi(\text{min})},\; \sigma_{\xi(\text{max})},\newline s_{\text{motion}}\;, \mathcal{G}$
\Ensure $n_{\text{max}}\geq 2,\;N_{\text{total}}\geq 100, \;s_{\text{motion}}\leq 10^{-3}$ 
\State \textbf{Sampling Modality of Noise $n_{\text{peaks}}$ and Computing}
\State  \textbf{Peak Weights $w$}
\State $n_{\text{peaks}} \sim U(2, n_{\text{max}})$
\State $w \gets U(0,1)\in\mathbb{R}^{n_{\text{peaks}}}$
\State $w \gets w \;/\;\| w \|$ \Comment{To ensure $\sum w = 1$}
\For{$peak$ in \textit{range}$(n_{\text{peaks}})$}
\State \textbf{Computing Gaussian Parameters for Each Peak}
\State $N_{\text{peak}} = N_{\text{total}} \times w[peak]$
\State $\mu_{\text{peak}} \sim U(\mu_{\xi(\text{min})}, \mu_{\xi(\text{max})}) \in \mathbb{R}^{o}$
\State $\sigma_{\text{peak}} \sim U(\sigma_{\xi(\text{min})}, \sigma_{\xi(\text{max})}) \in \mathbb{R}^{o}$ \\
\State \textbf{Sampling $N_{\text{peak}}$ Data Points from Defined Gaussian}
\State \textbf{and Storing Them into Array $\mathcal{G}$}
\State $g \sim N(\mu_{\text{peak}},\sigma_{\text{peak}}) \in \mathbb{R}^{N_{\text{peak}} \times o}$
\State $ \mathcal{G} \gets \mathcal{G}.\text{concatenate}(g)$ 
\EndFor
\State \textbf{Randomly-Selecting Single Noise Vector from }$\mathcal{G}\in\mathbb{R}^{N_{\text{total}}\;\times\; o}$
\State $\xi_k \gets \mathcal{G}.\text{sample}()\in\mathbb{R}^{o}$ \\
\State \textbf{Applying Sampled Noise Vector $\xi_k$ to Observed States $y_k$}
\State $y_{k} \gets y_{k} + \xi_k$
\State $y_{k} \gets y_{k} \times (1 + s_{\text{motion}}\;\times\; y_{k}) $
\State \textbf{return} $y_{k}$
\end{algorithmic}
\end{algorithm}

\bibliography{sn-article}  

\end{document}